\documentstyle[psfig]{mn}

\newcommand{\hmpc}{$h^{-1}$\,Mpc}
\newcommand{\hkpc}{$h^{-1}$\,Kpc}
\newcommand{\be}{\begin{equation}}
\newcommand{\e}{\end{equation}}
\newcommand{\f}{\frac}

\title{Semi analytic approach to understanding the distribution 
of neutral hydrogen in the universe}
\author
	[T. Roy Choudhury, T. Padmanabhan \& R. Srianand]
       {T. Roy Choudhury\thanks{E-mail: tirth@iucaa.ernet.in}, 
	T. Padmanabhan\thanks{E-mail: paddy@iucaa.ernet.in}, 
	R. Srianand\thanks{E-mail: anand@iucaa.ernet.in} \\
	IUCAA, Post Bag 4, Ganeshkhind, Pune 411 007, India.}
\date{Accepted 2000 December 00.
      Received 2000 December 00;
      in original form 2000 December 00}

\pagerange{\pageref{firstpage}--\pageref{lastpage}}
\pubyear{2000}

\begin{document}

\maketitle

\label{firstpage}

\begin{abstract}
Analytic derivations of the correlation
function and the column density distribution for neutral hydrogen in 
the intergalactic medium (IGM) are presented, 
assuming that the non-linear baryonic 
mass density distribution in the IGM is lognormal. This ansatz
was used earlier by Bi \& Davidsen (1997) to perform 1D
simulations of lines-of-sight and analyse the properties of
absorption systems. We have taken a completely analytic approach, which
allows us to explore a wide region of the parameter space for our model.  
The analytic results have been compared with 
observations to constrain various cosmological 
and IGM parameters, whenever possible.
Two kinds of correlation functions are defined : (i) along 
the line-of-sight (LOS) and (ii) across the transverse direction. 
We find that the effects on the LOS correlation 
due to change in cosmology and the slope of the 
equation of state of the IGM, $\gamma$ are 
of the same order, which means that we cannot constrain both the 
parameters simultaneously. However,
it is possible to constrain $\gamma$ and 
its evolution using the observed LOS correlation function at different 
epochs provided one knows the background cosmology. We suggest that 
the constraints on the evolution of 
$\gamma$ obtained using the LOS correlation 
can be used as an independent tool to probe the reionisation 
history of the universe. 
From the transverse correlation function, we obtain the excess probability, 
over random, of finding two neutral hydrogen overdense regions separated by an
angle $\theta$. We find that this excess probability is always less than 
1 per cent 
for redshifts greater than 2.
Our models also reproduce the observed column density distribution 
for neutral hydrogen and the shape of 
the distribution depends on $\gamma$. 
Our calculations suggest that one can rule out $\gamma > 1.6$ for 
$z \simeq 2.31$ using 
the column density distribution. However, one cannot rule higher values 
of $\gamma$ at higher redshifts.
\end{abstract}


\section{Introduction}
\label{intro}

The nature and evolution of the initial power spectrum of density
fluctuations could be obtained by studying the distribution of objects
at different scales and different epochs.
The formalism for
studying the formation of dark matter (DM) structures is well
established, as they are collisionless particles interacting only
through gravity, and has been extensively studied using the large
cosmological N-body simulations. However, in order to model the
evolution of baryonic structures like galaxies, groups of galaxies etc. one
needs to incorporate all the hydrodynamical processes, heating,
cooling, star formation etc., in the N-body simulations. Because of 
such complications, our understanding of the formation of baryonic structures 
has been limited.

Among the various baryonic structures, the regions where one can neglect the 
star formation are comparatively easier to study. Two such areas are 
(a) low amplitude fluctuations in the intergalactic medium (IGM), where 
the star formation rate is very low, and 
(b) the intracluster medium, where one studies processes 
over large scales and thus 
the star formation details can be neglected. 
Hence considerable effort has been given in understanding these two 
types of structures. 

The baryonic matter distribution at $z\le5$ is well probed through the
absorption signatures they produce on the spectra of the distant QSOs.
It is widely believed that while the metal lines systems (detected
through Mg~{\sc ii} or C~{\sc iv} doublets) seen in the QSO spectra
could be associated with the halos of the intervening luminous galaxies
(Bergeron \& Boisse 1991; Steidel 1993), most of the low neutral
hydrogen column density absorption lines (commonly called as
`Ly$\alpha$' clouds) are believed to be due to low amplitude baryonic
fluctuations in the IGM.

Semi analytical as well as hydrodynamical simulations are consistent 
with the view that the
Ly$\alpha$ clouds are small scale density fluctuations
(Bond, Szalay \& Silk 1988; Cen et al. 1994; Zhang, Anninos 
\& Norman 1995; Hernquist et al. 1996; Miralda-Escud\'e et al. 1996; 
Bi \& Davidsen 1997; Riediger, Petitjean \& M\"ucket 1998;  
Theuns, Leonard \& Efstathiou 1998; Theuns et al. 1998; 
Dav\'e et al. 1999) that are
naturally expected in any standard structure formation models. This
idea is supported by the detection and the evolution of the weak
clustering among the Ly$\alpha$ clouds in the redshift space
(Cristiani et al. 1995; Srianand 1997; Khare et al. 1997).
Subsequently it is realised that the thermal history of the Ly$\alpha$
line forming regions depends on (i) epoch of reionisation (equation of
state) (ii) rate of photoionisation and (iii) adiabatic cooling.  One
can in principle neglect shocks and other processes that are important only 
in the highly non-linear regime. However a simple linear evolution of
the densities will fail to produce the saturated Ly$\alpha$ systems
and one needs to incorporate non-linearities in the model.

As a first step,  one can model the non-linear evolution of the baryonic
fluctuations that produce Ly$\alpha$ clouds using one of the several 
approximations like: (i) Zeldovich
approximation (Doroshkevich \& Shandarin 1977; McGill 1990; Hui, Gnedin \& 
Zhang 1997), (ii) lognormal approximation
(Bi 1993; Gnedin \& Hui 1996; Bi \& Davidsen 1997). 
or (iii) power law
approximation (Bi, Ge \& Fang 1995) (strictly speaking, 
the baryonic fluctuations are calculated here using the linear theory). 
In all these cases the baryon
density is estimated from the DM density by some rule 
and the neutral fraction is
estimated by considering the equilibrium between the rate of
photoionisation due to background radiation and the rate of
recombination estimated from the temperature defined through the
equation of state. All these models depend on various IGM 
parameters such as intensity of the background radiation,
equation of state and  density-averaged temperature as well as
the cosmological parameters like $\Omega_m, \Omega_{\Lambda},$ etc.

Observationally the statistical properties of the Ly$\alpha$
absorption lines are quantified through the column density
distribution, correlation functions and their dependence on the mean redshift.
The clustering properties of the Ly$\alpha$ absorption lines are
studied through two point correlation function obtained either (a) in the
redshift space using the lines detected along a single line-of-sight (LOS)
which we call ``LOS correlation function'', or (b) among the
absorption lines detected along the lines of sights toward a few
closely spaced QSOs which we call ``transverse correlation function''.
In either case the observed spectra is decomposed into clouds using
``Voigt'' profile fits. Though this process smoothens the density field
over the width of the lines the average effects due to thermal
broadening is taken care of by the Voigt profiles. One can also compute 
the two-point correlation function of the observed flux in different
pixels. As this process does not decompose the actual density fields
into cloudlets, in order to analyse the data the models should
incorporate the thermal broadening and blending of contribution from
different density fluctuations (Croft et al. 1999; McDonald et al. 1999).  
Most of the existing studies concentrate on
obtaining constraints on the cosmological parameters using the observed
statistical properties.  Comparatively very less effort is directed 
to understand how
the observed quantities depend on the physical conditions in the IGM.

In this work we make a preliminary attempt  to investigate
the dependence of the observable quantities on  
various parameters of the models using a simple analytic approach.
We derive analytic relations
for the two-point correlation function among the Ly$\alpha$ clouds
and the column density distribution using a lognormal approximation.
These equations are used with the observed Voigt profile fitted data to get
constraints on different IGM and cosmological parameters.
In section \ref{analytic},
we treat the non-linear evolution with a simple ansatz proposed by Bi
\& Davidsen (1997) for the baryonic density fluctuations, and derive
analytic expressions for the correlation function along the LOS 
and in the transverse direction and the column density
distribution. The model parameters used to obtain various 
results are discussed in section \ref{mod_par}.
In section \ref{results}, we study the 
correlation function at different redshifts for different structure
formation models and for different values of the IGM parameters such 
as the density averaged temperature and the
equation of state. We compare some of our results with the existing
observational data.  
We also present the results for the column density 
distribution and study its dependence on various cosmological and IGM 
parameters. The results are summarised in section \ref{concl}.

\section{Analytic Model}
\label{analytic}

The linear density contrast for dark matter in comoving $k$-space, for a 
particular redshift $z$, is given by
\be
\delta_{\rm DM}(\mbox{\boldmath $k$},z)=
D(z) \delta_{\rm DM}(\mbox{\boldmath $k$},0),
\e
where $D(z)$ is the linear growth factor for the density contrast, normalised 
such that $D(0)=1$. 
If we assume the linear density contrast to be a Gaussian 
random field,
then the corresponding linearly extrapolated power spectrum 
$P_{\rm DM}(k)$ is defined by
\be
\langle\delta_{\rm DM}(\mbox{\boldmath $k$},0)
\delta_{\rm DM}(\mbox{\boldmath $k^{\prime}$},0)\rangle
=(2\pi)^{3}P_{\rm DM}(k)
\delta_{\rm{Dirac}}(\mbox{\boldmath $k$}-\mbox{\boldmath $k^{\prime}$}).
\e
The power spectrum is only a function of 
the magnitude of $\mbox{\boldmath $k$}$, because of the isotropy of the 
background universe.

The linear density contrast for baryons in the IGM can
be obtained from the DM density contrast by smoothing over scales below the
Jeans length. We use the relation (Fang et al. 1993)
\be
\delta_{\rm B}(\mbox{\boldmath $k$},z)=
\f{\delta_{\rm DM}(\mbox{\boldmath $k$},z)}{1+x_b^2(z)k^2},
\label{eq:delta_b}
\e
where
\be
x_b(z)=\f{1}{H_0}\left[\f{2\gamma k_B T_m(z)}{3\mu m_p \Omega_m (1+z)}\right]
^{1/2}
\label{eq:xb}
\e
is the Jeans length; $T_m$ and $\mu$ are the density-averaged temperature and 
mean molecular 
weight of the IGM respectively; $\Omega_m$ is the cosmological 
density parameter of total 
mass and $\gamma$ is the ratio of specific heats. Strictly speaking, 
equation (\ref{eq:delta_b}) is valid only for the case where $x_b$ is 
independent of $z$, but it 
is shown by Bi, Borner \& Chu (1992) that equation (\ref{eq:delta_b}) is a 
good approximation for 
$\delta_{\rm B}(\mbox{\boldmath $k$},z)$ 
even when $x_b$ has a redshift dependence.
The linear density contrast in real comoving space, 
$\delta(\mbox{\boldmath $x$},z)$, 
is the Fourier transform of
equation (\ref{eq:delta_b}).

In principle, to study the properties of the IGM one has to take into
account the non-linearities in the density distribution and various
physical processes such as shocks, radiation field, cooling
etc. However, detailed hydrodynamical modelling of IGM has shown that most
of the low column density Ly$\alpha$ absorption (i.e. $N_{\rm HI}\le10^{14}$
cm$^{-2}$) are produced by regions that are either in the linear 
or in the weakly non-linear regime (Cen et al. 1994; Zhang, Anninos 
\& Norman 1995; Hernquist et al. 1996; Miralda-Escud\'e et al. 1996; 
Theuns, Leonard \& Efstathiou 1998; Theuns et al. 1998; 
Dav\'e et al. 1999). The lower envelope of the column density, 
$N_{\rm HI}$ versus the thermal velocity dispersion, $b$ 
(given by $b=(2 k_B T/m_p)^{1/2}$) scatter plot  
(Schaye et al. 1999a; Schaye et al. 1999b) 
suggests that 
there is a well defined relationship between the density and the
temperature of the IGM (Hui \& Gnedin 1997). Thus it is
possible to model low column density systems using simple prescription for
the non-linear density field and an equation of state.

In this work, we take into account the effect of non-linearity by
assuming the number density distribution of the baryons, 
$n_{\rm B}(\mbox{\boldmath $x$},z)$ to be
a lognormal random field
\be
n_{\rm B}(\mbox{\boldmath $x$},z)=
A {\rm e}^{\delta_{\rm B}(\mbox{\boldmath $x$},z)}
\e
where $A$ is a constant to be determined. The mean value of 
$n_{\rm B}(\mbox{\boldmath $x$},z)$
is given by
\be
\langle n_{\rm B}(\mbox{\boldmath $x$},z) \rangle \equiv n_0(z)
= A \langle {\rm e}^{\delta_{\rm B}(\mbox{\boldmath $x$},z)} \rangle
\e
Since $\delta_{\rm B}(\mbox{\boldmath $x$},z)$ is a Gaussian random field, 
one can write
\be
\langle {\rm e}^{\delta_{\rm B}(\mbox{\boldmath $x$},z)} \rangle= 
{\rm e}^{\Delta^2(z)/2}
\label{eq:expdel}
\e
where
\be
\Delta^2(z)=\langle\delta_{\rm B}^2(\mbox{\boldmath $x$},z)\rangle=D^2(z)
\int\f{{\rm d}^3k}{(2\pi)^3} \f{P_{\rm DM}(k)}{(1+x_b^2(z)k^2)^2}.
\label{eq:deltasq}
\e
Hence,
\be
A=n_0(z) {\rm e}^{-\Delta^2(z)/2}
\e
and
\be 
n_{\rm B}(\mbox{\boldmath $x$},z)=n_0(z) 
\exp [\delta_{\rm B}(\mbox{\boldmath $x$},z)-\f{\Delta^2(z)}{2}].
\label{eq:ln}
\e
The lognormal distribution was introduced by 
Coles \& Jones (1991) as a model for the non-linear matter distribution in the 
universe. This ansatz has several interesting features:\\
(a) 
It can be seen that the matter density given by 
equation (\ref{eq:ln}) is always positive, even when 
$\delta_{\rm B} \to -\infty$, unlike any polynomial function of 
$\delta_{\rm B}$.
When the
density contrast is small ($\delta_{\rm B} \ll 1$), 
equation (\ref{eq:ln}) reduces 
to 
$n_{\rm B}/n_0 \simeq 1+\delta_{\rm B}$, which is just what 
we expect from linear theory.\\
(b) On small 
scales, equation (\ref{eq:ln}) becomes the isothermal hydrostatic solution, 
which 
describes highly clumped structures like intracluster gas, 
$n_{\rm B} \propto \exp 
(-\mu m_p \psi_{\rm DM} /\gamma k_B T)$, where 
$\psi_{\rm DM}$ is the dark matter
potential (Sarazin \& Bahcall 1977). The lognormal function can be thought 
of as the simplest function which links these two extreme regions smoothly.\\
(c) One can also think of 
the lognormal distribution as the kinematic model for the 
density field.
If one assumes that the initial density and velocity fields are Gaussian,  
and extrapolates the continuity equation into non-linear regimes,
treating the velocity field as linear,  
it turns out that the non-linear density field obtained in such a manner 
follows the lognormal distribution (Coles \& Jones 1991).\\
(d) Bi \& Davidsen (1997) have tested the distribution against hydrodynamical 
simulations, and found a reasonable match between them. 
The lognormal assumption
has also been used to model the IGM in numerical simulations (Bi 1993; Bi \&
Davidsen 1997) and is found to be working well 
in reproducing the observations. In
particular, the simulation results matches well with the observed column 
density 
distribution and number density of the
Ly$\alpha$ absorption lines, the probability distribution of the $b$ 
parameter etc (see Bi \& Davidsen 1997). 

We shall also discuss briefly a more general argument as to 
why the lognormal distribution 
should be natural choice in a large class of phenomena. 
There is a wide class of quantities, denoted by $f$, 
the time evolution of which can be characterised by the 
following property -- the change in the value of $f$ at some instant $t_i$ is  
proportional to its value at that instant, with the proportionality factor 
being a random variable. In mathematical notation, this can be written as 
$f(t_{i+1})=f(t_i)+\varepsilon_i f(t_i)$, where $\varepsilon_i$ 
is the random variable. (Some examples of such phenomena in 
sociological context are (i) rich getting 
richer through fluctuations in stock market, and (ii) more facilities 
being provided to people who already have them.) Similar situation can 
occur in structure formation scenario also. The regions which 
have high density, because of stronger gravitational attraction, 
have a better chance of acquiring more mass. 
Let us denote 
the density field at some particular point at a given epoch $t_i$ by $n(t_i)$ 
and postulate the evolution, 
\be
n(t_{i+1})=n(t_i)+\varepsilon_i n(t_i)=(1+\varepsilon_i) n(t_i)
\e
We can now write $n(t_{i+1})$ in terms of 
some initial density field $n(t_0)$
\be
n(t_{i+1})=(1+\varepsilon_i)(1+\varepsilon_{i-1}) 
\dots (1+\varepsilon_0)n(t_0)
\e
Taking logarithm of both sides
\be
\ln[n(t_{i+1})]=\sum_{j=0}^i \ln (1+\varepsilon_j)+\ln[n(t_0)]
\e
It is clear that when the time intervals $(t_{i+1}-t_i)$ are small, the mass 
acquired within that interval will also be very small. 
Hence we expect that $\varepsilon_i \ll 1$. 
Then the above expression becomes 
\be
\ln\left[\f{n(t_{i+1})}{n(t_0)}\right]=\sum_{j=0}^i \varepsilon_j
\e
This means that $\ln[n(t)/n(t_0)]$ is a sum of a large number of uncorrelated 
random variables. Using the central limit theorem, we can conclude that it 
follows a Gaussian distribution or, 
equivalently, $n(t)$ follows a lognormal distribution. This suggests that 
it may be  
reasonable to try an ansatz that the distribution of the 
non-linear baryonic density field is lognormal.

As an aside, we just mention that our 
analysis described here can easily be carried out 
for any other local 
ansatz for the non-linear baryonic density.
[The results for power law 
assumption in which 
$n_{\rm B} \propto (1+\delta)^{p}$ will be discussed in a later 
paper.]

Once we have obtained the total baryonic density, 
the fraction of hydrogen in the neutral form, $f$,
in the IGM can be obtained by solving the ionisation
equilibrium equation
\be
\alpha(z,T(z)) n_p n_e=J(z) n_{\rm HI},
\label{eq:ion_eq}
\e
where $\alpha(z,T(z))$ is the radiative recombination rate and $J(z)$ 
is rate of photoionisation for hydrogen at redshift $z$ (Black 1981); $n_p, 
n_e$ and
$n_{\rm HI}$ are the number densities of proton, 
electron and neutral hydrogen, 
respectively.
For simplicity, we assume that hydrogen is the only element present in 
the IGM and 
neglect the presence of helium and other heavier elements. 
In such a case, we 
have $n_e=n_p$. (This relation is not valid in the presence of helium or 
other heavier elements. If we have taken their 
presence into account, we would have got $n_e= \kappa n_p$, 
where $\kappa$ is a constant. Usually, $1 \le \kappa \le 1.2$, 
because the amount 
of helium and heavier elements in the IGM is 
small compared to hydrogen. Since we do not know $J(z)$ beyond 
an accuracy of 10--20 per cent, we can always absorb $\kappa$ into 
$J(z)$.) Let us define the neutral fraction of hydrogen, 
$f$ by
\be
f=\f{n_{\rm HI}}{n_{\rm B}}=\f{n_{\rm HI}}{n_{\rm HI}+n_p} 
\e
Hence we get from equation (\ref{eq:ion_eq})
\be
\f{(1-f)^2}{f}=\f{J(z)}{\alpha(z,T(z)) n_{\rm B}}.
\e
In general, one can solve this equation and determine $f$ as a function 
of $n_{\rm B}$.
This expression simplifies for two extreme cases. For $f \ll 1$, we get
\be
f=\f{\alpha(z,T(z)) n_{\rm B}}{J(z)}
\e
and for $f \sim 1$,
\be
f=1-\sqrt{\f{J(z)}{\alpha(z,T(z)) n_{\rm B}}}.
\e
Hence, we have
\be
n_{\rm HI}(\mbox{\boldmath $x$},z)=\left\{ \begin{array}{ll}
	\!\!\f{\alpha(z,T(z))}{J(z)} n_{\rm B}^2(\mbox{\boldmath $x$},z) 
		&\!\!\!\!\!\!\mbox{(if $n_{\rm HI} \ll n_{\rm B}$)}\\
	\!\!\!n_{\rm B}(\mbox{\boldmath $x$},z)-\!
	\sqrt{\f{J(z) n_{\rm B}(\mbox{\boldmath $x$},z)}{\alpha(z,T(z))}} 
			&\!\!\!\!\!\mbox{(if $n_{\rm HI} \sim n_{\rm B}$)}
               \end{array} \right.
\label{eq:nh(n)}
\e
The ionisation conditions in the Ly$\alpha$ absorbers are similar
to the of H~{\sc ii} regions with $f\simeq 10^{-4}$. Thus, from now on
we concentrate only on the case $n_{\rm HI}\ll n_{\rm B}$.

We take the temperature dependence of the recombination coefficient $\alpha$ 
to be given by (Rauch et al. 1997)
\be
\alpha(z,T(z))=\alpha_0 \left(\f{T(z)}{10^4 K}\right)^{-0.7},
\e
where $\alpha_0=4.2 \times 10^{-13}$ cm$^3$ s$^{-1}$. This relation is a good 
approximation for $\alpha$ in the temperature range relevant for Ly$\alpha$ 
forest. 
The temperature $T$ is related to the baryonic density $n$ through the 
equation of state. We assume a polytropic equation of state 
$p \propto \rho^{\gamma} \propto n^{\gamma}$, or equivalently 
\be
T(z) = T_0(z) [n_{\rm B}(z)/n_0(z)]^{\gamma-1},
\e
where 
\be
n_0(z)=\f{\Omega_{\rm baryon} \rho_c}{\mu_b m_p} (1+z)^3
\label{eq:n0}
\e
is the mean baryonic 
number density at redshift $z$. $\rho_c$ is the critical matter density at
the present epoch, given by
\be
\rho_c=1.8791 \times 10^{-29} h^2 {\rm cm}^{-3}
\e
and $\mu_b m_p$ is the mass per baryonic particle.
Then, the H~{\sc i} density becomes
\be
n_{\rm HI}(\mbox{\boldmath $x$},z)=
F(z) \left(\f{n_{\rm B}(\mbox{\boldmath $x$},z)}{n_0(z)}\right)^{\beta}
\label{eq:nh_nb_beta}
\e
where
\be
F(z)=\alpha_0 n_0^2(z) \left(\f{T_0(z)}{10^4 K}\right)^{-0.7} J^{-1}(z)
\label{eq:fz}
\e
and
\be
\beta=2.7-0.7 \gamma.
\label{eq:beta}
\e
(We note, in passing, that 
$\beta$ becomes negative if $\gamma > 3.86$.) 
We can write the H~{\sc i} density in terms of the linear baryonic density 
contrast
\be 
n_{\rm HI}(\mbox{\boldmath $x$},z)=
F_1(z)\exp[\beta \delta_{\rm B}(\mbox{\boldmath $x$},z)],
\label{eq:nh}
\e
where
\be
F_1(z)=F(z) {\rm e}^{-\beta \Delta^2(z)/2}.
\e
It is clear from equation (\ref{eq:nh}) that the H~{\sc i} distribution at a
particular redshift is also described by a lognormal distribution. All the
statistical quantities regarding H~{\sc i} can be derived from this in a
straightforward manner. 

\subsection{Correlation Function for Neutral Hydrogen}
\label{sub_xi}

One of our main interest is the correlation function 
\begin{eqnarray} 
\lefteqn{ 
\langle
n_{\rm HI}(\mbox{\boldmath $x$},z) 
n_{\rm HI}(\mbox{\boldmath $x^{\prime}$},z^{\prime})\rangle= 
}\nonumber \\
& & F_1(z) F_1(z^{\prime}) \langle \exp\{\beta 
[\delta_{\rm B}(\mbox{\boldmath $x$},z) +
\delta_{\rm B}(\mbox{\boldmath $x^{\prime}$},z^{\prime})]\}\rangle,
\label{eq:nhnh}
\end{eqnarray}
from which several useful quantities can be obtained. 
Since $\delta_{\rm B}$ is a Gaussian random field, 
we can write, using equation 
(\ref{eq:expdel})
\begin{eqnarray}
\lefteqn{
\langle 
\exp \{\beta [\delta_{\rm B}(\mbox{\boldmath $x$},z) 
+\delta_{\rm B}(\mbox{\boldmath $x^{\prime}$},z^{\prime})]\}
\rangle =
}\nonumber \\
& &\exp[\f{\beta^2}{2} \{\Delta^2(z) + \Delta^2(z^{\prime}) 
+2 Q(\mbox{\boldmath $x$},\mbox{\boldmath $x^{\prime}$};z,z^{\prime})\}],
\end{eqnarray}
where
\be
Q(\mbox{\boldmath $x$},\mbox{\boldmath $x^{\prime}$};z,z^{\prime})=
\langle \delta_{\rm B}(\mbox{\boldmath $x$},z)
\delta_{\rm B}(\mbox{\boldmath $x^{\prime}$},z^{\prime}) \rangle.
\e
Simple algebra gives
\begin{eqnarray}
\lefteqn{
Q(\mbox{\boldmath $x$},\mbox{\boldmath $x^{\prime}$};z,z^{\prime}) \equiv 
Q(\mbox{\boldmath $x$}-\mbox{\boldmath $x^{\prime}$};z,z^{\prime})=
} \nonumber \\
& & D(z) D(z^{\prime}) \int \f{{\rm d}^3k}{(2\pi)^3} \f{P_{\rm DM}(k) 
{\rm e}^{{\rm i}\mbox{\boldmath $k$}\cdot(\mbox{\boldmath $x$}-
\mbox{\boldmath $x^{\prime}$})}}
{(1+x_b^2(z)k^2)(1+x_b^2(z^{\prime})k^2)}.
\label{eq:q} 
\end{eqnarray}
One also notes from equations (\ref{eq:deltasq}) and (\ref{eq:q}) that
$\Delta^2(z)=Q(0;z,z)$. We can now write equation  (\ref{eq:nhnh}) as
\be
\langle n_{\rm HI}(\mbox{\boldmath $x$},z) 
n_{\rm HI}(\mbox{\boldmath $x^{\prime}$},z^{\prime})\rangle=
F_2(z) F_2(z^{\prime}) 
{\rm e}^{\beta^2 Q(\mbox{\boldmath $x$}-
\mbox{\boldmath $x^{\prime}$};z,z^{\prime})},
\e
where
\be
F_2(z)=F_1(z){\rm e}^{\beta^2 \Delta^2(z)/2}.
\e

One needs to normalise the quantity \linebreak 
$\langle n_{\rm HI}(\mbox{\boldmath $x$},z) 
n_{\rm HI}(\mbox{\boldmath $x^{\prime}$},z^{\prime})\rangle$, 
to obtain the correlation 
function $\xi_{\rm HI}(\mbox{\boldmath $x$}-
\mbox{\boldmath $x^{\prime}$};z,z^{\prime})$ 
for H~{\sc i}. 
A natural 
way of normalising the correlation would be to use the definition
\be
1+\xi_{\rm HI}(\mbox{\boldmath $x$}-
\mbox{\boldmath $x^{\prime}$};z,z^{\prime})=
\f{\langle n_{\rm HI}(\mbox{\boldmath $x$},z) 
n_{\rm HI}(\mbox{\boldmath $x^{\prime}$},z^{\prime})\rangle}
  {\langle n_{\rm HI}(\mbox{\boldmath $x$},z)\rangle
   \langle n_{\rm HI}(\mbox{\boldmath $x^{\prime}$},z^{\prime})\rangle}.
\e
Since $\langle n_{\rm HI}(\mbox{\boldmath $x$},z)\rangle=F_2(z)$, we get
\be
\xi_{\rm HI}(\mbox{\boldmath $x$}-\mbox{\boldmath $x^{\prime}$};z,z^{\prime})=
{\rm e}^{\beta^2 Q(\mbox{\boldmath $x$}-
\mbox{\boldmath $x^{\prime}$};z,z^{\prime})}-1
\label{eq:xi_q}
\e
with $Q$ given by equation (\ref{eq:q}).

All the analysis above is valid if one can probe any scale with arbitrary
accuracy. But it turns out that one cannot obtain information about scales
smaller than some particular value, due to various observational constraints.
While observing along a LOS, it will be impossible to 
probe the velocity scales less than the spectroscopic limit due to
thermal broadening and the blending of spectral lines. Similarly,
while observing across the transverse direction, the peculiar velocities of
individual points will constrain the velocity resolution which we have
not taken into account in the above
analysis. If $\Delta v$ is the smallest scale one can probe, then
the corresponding limit in the redshift-space is
\be 
\Delta z=\f{\Delta v}{c} (1+z). 
\e 
This means that we will not be able to probe
below a comoving length scale given  by  
\be
\Delta x(z)=d_H(z) \Delta z,
\label{eq:delta_x}
\e
where
\begin{eqnarray}
d_H(z)&=&c\left(\f{\dot{a}}{a}\right)^{-1} \nonumber \\
      &=&\f{c}{H_0} [\Omega_{\Lambda}+\Omega_m (1+z)^3+\Omega_k
(1+z)^2]^{-1/2}, 
\label{eq:d_h}
\end{eqnarray}
\be
\Omega_k=1-\Omega_m-\Omega_{\Lambda}.
\label{eq:om_k}
\e
This effect can be included in our calculation by smoothing over all the 
length 
scales smaller than $\Delta x(\bar{z})$, where 
\be
\bar{z}=\f{1}{2}(z + z^{\prime})
\label{eq:z_bar}
\e
is the average redshift. We use a Gaussian window of width 
$\sigma _x(\bar{z})=
\Delta x(\bar{z})$, and get a smoothed version of $Q$ in 
equation (\ref{eq:q}).
In Fourier space, this smoothing will introduce an extra Gaussian term in the 
integrand, and our smoothed $Q$ will be
\begin{eqnarray}
\lefteqn{
Q_{\rm smooth}(\mbox{\boldmath $x$}-
\mbox{\boldmath $x^{\prime}$};z,z^{\prime})=
} \nonumber \\
& & D(z) D(z^{\prime}) \int \f{{\rm d}^3k}{(2\pi)^3} \f{P_{\rm DM}(k) 
{\rm e}^{-k^2 \sigma_x^2(\bar{z})/2} 
{\rm e}^{{\rm i}\mbox{\boldmath $k$}\cdot(\mbox{\boldmath $x$}
-\mbox{\boldmath $x^{\prime}$})}}
{(1+x_b^2(z)k^2)(1+x_b^2(z^{\prime})k^2)}. 
\end{eqnarray}
The angular integrations can be carried out trivially, and we get
\begin{eqnarray}
\lefteqn{
Q_{\rm smooth}(\mbox{\boldmath $x$}-
\mbox{\boldmath $x^{\prime}$};z,z^{\prime})=\f{D(z) D(z^{\prime})}{2 \pi^2} 
\times
} \nonumber \\
& & 
\int_0^{\infty} {\rm d}k 
\f{P_{\rm DM}(k) k^2{\rm e}^{-k^2 \sigma_x^2(\bar{z})/2}} 
{(1+x_b^2(z)k^2)(1+x_b^2(z^{\prime})k^2)}
\f{\sin kX}{kX},
\end{eqnarray}
where $X=|\mbox{\boldmath $x$}-\mbox{\boldmath $x^{\prime}$}|$.
The final integration can be done numerically, once the DM power spectrum is 
given.

At this stage, the relations derived above can be used for any 
$\mbox{\boldmath $x$},
\mbox{\boldmath $x^{\prime}$}, z, z^{\prime}$. 
As we mentioned earlier, if one observes the H~{\sc i} along a particular 
LOS, then one is probing  
different regions of the IGM at different redshifts. The position $x$ will be 
related to the redshift $z$ by the relation 
\be
x(z)=\int_0^z d_H(z^{\prime}) {\rm d}z^{\prime}
\label{eq:x(z)}
\e
where $d_H(z)$ is given by equation (\ref{eq:d_h}).
Then the LOS correlation function is given by 
\be
\xi_{\rm HI}^{\rm LOS}(z,z^{\prime})=
{\rm e}^{\beta^2 Q_{\rm LOS}(l(z,z^{\prime});z,z^{\prime})}-1,
\e
where
\begin{eqnarray}
\lefteqn{
Q_{\rm LOS}(l(z,z^{\prime});z,z^{\prime})=
} \nonumber \\
& & \f{D(z) D(z^{\prime})}{2 \pi^2} \int_0^{\infty} {\rm d}k 
\f{P_{\rm DM}(k) k^2{\rm e}^{-k^2 \sigma_x^2(\bar{z})/2}} 
{(1+x_b^2(z)k^2)(1+x_b^2(z^{\prime})k^2)}
\f{\sin kl}{kl},\nonumber\\
& &
\label{eq:q_los}
\end{eqnarray}
and
\be
l(z,z^{\prime})=x(z)-x(z^{\prime}).
\label{eq:l}
\e
It should be stressed that $\xi_{\rm HI}^{\rm LOS}(z,z^{\prime}) \neq 
\xi_{\rm HI}^{\rm LOS}(z-z^{\prime})$. 
This means that one cannot rigorously define a power 
spectrum from the LOS correlation function because the correlation is a 
function of {\em two} variables $z$ and $z^{\prime}$. {\sl In other words, 
the LOS 
power 
spectrum does not exist in strict sense.} However, one can get an approximate 
LOS power spectrum for a small redshift range around any mean redshift. 
We have already defined the average redshift in equation (\ref{eq:z_bar}).
We define a redshift difference 
\be
\Delta z= z - z^{\prime}
\label{eq:z_dif}
\e
and evaluate the correlation function 
for a particular value of $\bar{z}$ as a 
function of 
$\Delta z$, i.e.,
\be
\xi_{\rm HI}^{\rm LOS}(\bar{z},\Delta z)=
{\rm e}^{\beta^2 Q_{\rm LOS}(\bar{z},\Delta z)}-1,
\e
where
\begin{eqnarray}
\lefteqn{
Q_{\rm LOS}(\bar{z},\Delta z)=
} \nonumber \\
& & \f{D(\bar{z}+\Delta z/2) D(\bar{z}-\Delta z/2)}{2 \pi^2} 
\int_0^{\infty} {\rm d}k 
\left\{\f{\sin (k d_H(\bar{z}) \Delta z)}{k d_H(\bar{z}) \Delta z} \right.
 \nonumber\\
& &\left .\times\f{P_{\rm DM}(k) k^2{\rm e}^{-k^2 \sigma_x^2(\bar{z})/2}} 
{(1+x_b^2(\bar{z}+\Delta z/2)k^2)(1+x_b^2(\bar{z}-\Delta z/2)k^2)}\right\}
\nonumber\\
& &
\label{eq:q_los_delz}
\end{eqnarray}
For small $\Delta z$, one can use equation (\ref{eq:delta_x}) to 
write the correlation as a function of $\Delta x$, Fourier 
transform the correlation, and get the power 
spectrum. Such a power spectrum will depend on the value of $\bar{z}$. We 
stress again this power spectrum is approximate in the sense that it exists 
only for $\Delta z \ll \bar{z}$.

The transverse correlation is observed at some particular redshift 
($z=z^{\prime}$), along the transverse direction. Then
\be
\xi_{\rm HI}^{\rm trans}(l_{\perp};z)=
{\rm e}^{\beta^2 Q_{\rm trans}(l_{\perp};z)}-1,
\e
\begin{eqnarray}
\lefteqn{
Q_{\rm trans}(l_{\perp};z)=
} \nonumber \\
& & \f{D^2(z)}{2 \pi^2} \int_0^{\infty} {\rm d}k \f{P_{\rm DM}(k) k^2
{\rm e}^{-k^2 \sigma_x^2(z)/2}}{(1+x_b^2(z)k^2)^2}
\f{\sin kl_{\perp}}{kl_{\perp}},\nonumber\\
& &
\label{eq:q_sp}
\end{eqnarray}
where $l_{\perp}$ is the comoving distance along the transverse direction. For 
a given redshift, the 
transverse correlation is only a function of $l_{\perp}$. Hence, 
one can obtain the 
power spectrum from $\xi_{\rm HI}^{\rm trans}$ following usual methods.

\subsection{Column Density Distribution}
\label{sub_nh}

One of the other statistics the observers use to quantify the distribution 
Ly$\alpha$ absorption lines is column density distribution.
Indeed one can get the analytic expression for this using the
formalism developed so far in this work.
Note that Voigt profile fitting to the absorption lines are used to 
get the observed column density distribution.
Here we use a method called  `density-peak ansatz' (DPA), discussed in  
Gnedin \& Hui (1996) and Hui, Gnedin \& Zhang (1997) 
to derive an analytic expression 
for the column density distribution. 

Suppose we are looking at the IGM along any one direction, at some redshift 
$z$. Then the linear  density field $\delta_{\rm B}^{(1D)}(x,z)$ along that 
LOS will be described by a  one dimensional Gaussian random field.
DPA assumes that each density peak in the comoving space 
is associated with an absorption line, and one can assign a definite 
column density to each of them. In the articles referred above, 
each density peak is fitted with a Gaussian, and the column density 
is calculated using 
\be
N_{\rm HI} \propto \int_{\rm peak} n_{\rm HI}(x)~{\rm d}x.
\label{eq:n_prop}
\e 
In such a case, there is a definite 
correlation between the value of the density field at the peak, and the 
effective width of the absorber (which is determined by the correlation 
between the density field and its second derivative at the peak, and is fixed 
once the fitting function for the density peak is given).
We, however, take a simpler approach in assigning the column 
density to a density peak, which is described below.
 
The coherence scale of the distribution is defined as (Bardeen et al. 1986)
\be
R^* \equiv \f{\sigma_1}{\sigma_2},
\e
where $\sigma_1$ and $\sigma_2$ are defined in equation (\ref{eq:sigma_m}) 
(see Appendix A). 
This length is a measure of the distance between two successive zeroes for 
the one dimensional Gaussian random field. Since this is the 
relevant scale for the distribution of zero-crossing, we expect the 
effective length scale of a peak to be a fraction of $R^*$. Then  
the column density corresponding to a particular peak will be 
\be
N_{\rm HI} \propto 
n_{\rm HI}[\rm{peak}] R^* = n_{\rm HI}[\rm{peak}] R^* \epsilon
\label{eq:col_den}
\e
where $n_{\rm HI}$[peak] is the 
H~{\sc i} number density at the peak and $\epsilon$ is the proportionality 
constant, which can be used as a free parameter 
in comparing with observations. 
We have assumed $\epsilon$ to be independent of $N_{\rm HI}$, which means
that the column density is directly proportional to the peak density. 
Using this prescription for obtaining the column density from 
the H~{\sc i} density, we can easily obtain the 
relation between $N_{\rm HI}$ and the total baryonic over-density 
$n_{\rm B}/n_0$ using 
equation (\ref{eq:nh(n)}). For the case 
$n_{\rm HI}\ll n_{\rm B}$, the relation is given by
\be
\f{n_{\rm B}[\rm{peak}]}{n_0}=
\left(\f{N_{\rm HI}}{R^*\epsilon F(z)}\right)^{1/\beta}.
\label{eq:n_b_nh}
\e
The relation between $N_{\rm HI}$ and 
$\delta_{\rm B}^{(1D)}$, is then given by
\be
\delta_{\rm B}^{(1D)}[\rm{peak}]=\f{1}{\beta} \ln 
\left(\f{N_{\rm HI}}{R^*\epsilon F(z)}\right) 
+ \f{\Delta^2}{2}
\label{eq:delta_b_nh}
\e
Given this relation, it is straightforward to obtain the 
quantity ${\rm d}N_{\rm{pk}}/({\rm d}z~ {\rm d}N_{\rm HI})$, 
defined as the number of 
clouds (peaks) per column density interval per redshift interval. 
For completeness, we 
give the relevant calculations in the appendix.

\section{Model Parameters}
\label{mod_par}

In this section we discuss about the various model parameters used 
in obtaining the results. 
The parameters defining the model can be divided into two categories : (i) 
cosmological parameters, and (ii) parameters related to the IGM.

The first set of parameters are those which determine the 
background 
cosmology. We assume that the background universe is described by the FRW 
metric. We have considered four different cosmological models with the  
parameters listed below: 
\begin{enumerate}
\item[SCDM]
$\Omega_m=1, \Omega_{\Lambda}=0, h=0.5$
\item[OCDM]
$\Omega_m=0.35, \Omega_{\Lambda}=0, h=0.5$
\item[LCDM1]
$\Omega_m=0.35, \Omega_{\Lambda}=0.65, h=0.5$
\item[LCDM2]
$\Omega_m=0.35, \Omega_{\Lambda}=0.65, h=0.65$
\end{enumerate}
The next cosmological input that is required is the form of the DM power 
spectrum. We  take the following form for $P_{\rm DM}(k)$ 
(Efstathiou, Bond \& White 1992)
\be
P_{\rm DM}(k)=\f{A k}{(1 + [a k + (b k)^{1.5} + (c k)^{2}]^{\nu})^{2/\nu}}
\label{eq:ebw}
\e
where $\nu=1.13$, 
$a=(6.4/\Gamma)$\hmpc, $b=(3.0/\Gamma)$\hmpc, $c=(1.7/\Gamma)$\hmpc\ and 
$\Gamma=\Omega_m h$. The 
normalisation parameter $A$ is fixed through the value of $\sigma_8$ (the 
rms density fluctuation in spheres of radius 8 \hmpc).We take the values of 
$\sigma_8$ to be given by (Eke, Cole \& Frenk 1996)
\be
\sigma_8=\left\{ \begin{array}{ll}
	      (0.52 \pm 0.04) \Omega_m^{-0.46+0.10 \Omega_m} 
			&\!\!\!\!\!\! \mbox{(if $\Omega_{\Lambda}=0$)}\\
	      (0.52 \pm 0.04) \Omega_m^{-0.52+0.13 \Omega_m}
			&\!\!\!\!\!\! \mbox{(if $\Omega_{\Lambda}=1-\Omega_m$)}
		\end{array} \right.
\e

The next set of parameters are related to the physical conditions in the IGM.  
The parameters we need to describe the IGM are $\gamma$, $T_m$, 
$\Omega_{\rm baryon}$, $J(z)$ and $T_0(z)$. 
All these quantities were defined in section \ref{analytic}. 
Besides these, we have also introduced a 
parameter $\epsilon$ while modelling 
the column density distribution of the IGM. 
This is taken as a free parameter, to be fixed 
through observations.

It is known that
the value of $\gamma$, at any given epoch, depends on the reionisation history
of the universe (Hui \& Gnedin 1997). The value of $\gamma$ and its evolution 
is still quite uncertain.  Using
Voigt profile fits to the observed Ly$\alpha$ absorption lines one can in
principle obtain the value of $\gamma$.  Available observations are consistent
with $\gamma$ in the range $1.2 < \gamma < 1.7$ (Schaye et al.  1999b) 
for $2\le z \le 4.5$. As far as the evolution of $\gamma$ is concerned, we 
shall treat it as independent of $z$. 
The density averaged temperature is defined as,
\be
T_m=\f{\int \rho(T)~T~{\rm d}T}{\int \rho(T)~{\rm d}T}
\e
Using the equation of state $T \sim \rho^{\gamma -1}$, we get
\be
T_m=\f{2 \gamma-1}{\gamma} (T_{\rm max}-T_{\rm min}).
\e
We take $T_m$ to be in the range 10,000 K$<T_m<$60,000 K. The minimum value 
of $T_m$ 
corresponds to the minimum temperature of the IGM, which is determined 
by the photoionisation equilibrium. The maximum value of $T_m$  
corresponds to $b$ parameter $\sim 31.7$ km s$^{-1}$
and is consistent with the minimum value of $b$ observed at 
higher H~{\sc i} column densities (i.e. $10^{14.5}$ cm$^{-2}$). The evolution 
of $T_m$ depends on how $T_{\rm max}$ and $T_{\rm min}$ evolve with redshift. 
One possibility is to take the adiabaticity relation 
$T_m \sim (1+z)^{3\gamma -3}$.
However, people have argued that since there is no conclusive 
evidence of the evolution 
of the temperature in IGM, one should treat the mean temperature of the IGM 
as constant (Bi et al. 1995). Hence, in 
this paper, we also consider the second possibility where $T_m$ is independent 
of $z$. For the cases where we consider a small redshift range $\Delta z$ 
around a mean redshift $\bar{z}$, 
the effect of the evolution of $T_m$ is not very 
significant. But, whenever we study the evolution of a quantity over a 
large redshift range, we have to take into account the various 
redshift dependences of $T_m$.

We note that when we normalise the correlation function for 
neutral hydrogen, the parameters $\Omega_{\rm baryon}$, $J(z)$ and 
$T_0(z)$ cancel out. Hence the knowledge of these parameters are not necessary 
for modelling 
the correlation function. However, in the case of column density distribution, 
they appear as a combination 
$(\Omega_{\rm baryon}h^2)^2 J^{-1}(z) T_0^{-0.7}(z)$ through $F(z)$ 
(equations (\ref{eq:fz}) and (\ref{eq:n0})).
The values of  
these quantities are not known accurately, nor do we know how $J(z)$ and 
$T_0(z)$ evolve. In this work, we take $J(z)$ to be independent of $z$. We 
fix the value of the combination $(\Omega_{\rm baryon}h^2)^2 J^{-1}$ to be 
$(0.026)^2 \times 10^{12}$ s, which is consistent with the values given in 
McDonald et al. (1999). One should note that any change in the values of 
the above parameters can be compensated (to some extent) by changing the value 
of $\epsilon$, which is a free parameter in our model for the 
column density distribution.
For $T_0$ evolution, we shall consider two separate cases like  
$T_m$, i.e., (i) $T_0 \sim (1+z)^{3 \gamma -3}$ and (ii) $T_0$  
independent of $z$.

As we have discussed earlier, we need to smooth the power spectrum below some 
velocity because the blending of spectral lines makes it impossible to resolve 
the
lines below a particular velocity. Typically this velocity is of the order of
a few tens km s$^{-1}$. For definiteness, we take the smoothing velocity to be
$\Delta v=30$ km s$^{-1}$.

\section[]{Results}
\label{results}

\subsection{LOS Correlation}
\label{sub_LOS}

We shall now compute the results for the H~{\sc i} correlation 
function along the LOS as a function of the velocity separation $v$, where 
$v$ is related to $\Delta z$ by 
$v=c\Delta z/(1+\bar{z})$.

\begin{figure}
\psfig{figure=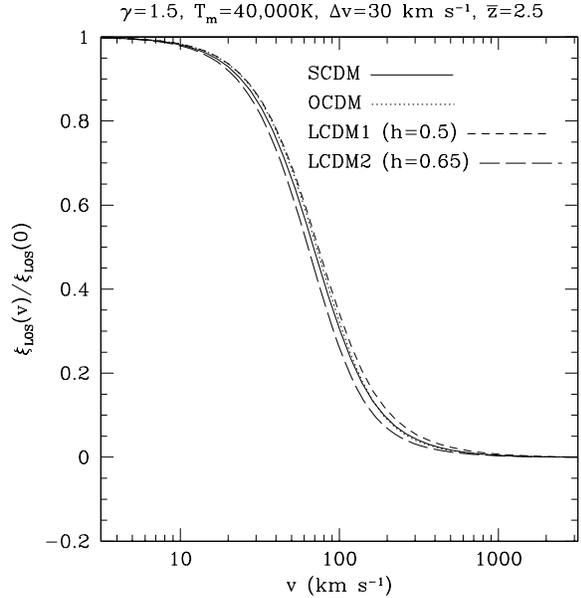,width=8cm}
\caption[]{\label{los}
LOS correlation function as a function of velocity separation. Results for 
four different cosmological models at a mean redshift of $\bar{z}=2.5$ are 
presented. The 
correlation function is normalised to unity at zero 
velocity separation.}
\end{figure}

The results for the LOS correlation function for different 
cosmological models are shown in Figure \ref{los}.
We have chosen typical values for $T_m$ as 40,000 K and $\gamma=1.5$, at a 
redshift of $\bar{z}=2.5$. 
It can be seen that the correlation curves tend to flatten at low velocities, 
and goes to zero at high velocities.

Miralda-Escud\'e et al. (1996) give the correlation function of the 
transmitted flux along a LOS
(see the solid curve in their Figure 12a). We note that 
the correlation curves for the transmitted flux and the neutral hydrogen 
density need not be exactly the same. However, we expect that the broad 
features and the general trends should be alike. It turns out that 
the correlation curve for transmitted flux obtained from the simulations
does have the same trend as our results.

We can compare how the shape of $\xi_{\rm LOS}(v)$ depends on 
various parameters. 
We found that the $\xi_{\rm LOS}(v)$ curve 
falls approximately like a power law, 
$\xi_{\rm LOS}(v) \propto v^{-p}$, at velocities 
within 100--1000 km s$^{-1}$.
At lower velocities the curve is practically independent of $v$. We have 
given the value of $p$ and the rms error on $p$ 
for different cosmological models and for different 
IGM parameters in Table \ref{los_v}.

\begin{table}
\caption{Power law index, $p$,  
for LOS correlation function, defined by 
$\xi_{\rm LOS}(v) \propto v^{-p}$ in the velocity range 
100--1000 km s$^{-1}$ for $z=2.5$. 
We also give the $1 \sigma$ error in $p$.}
\begin{tabular}{|l|l|c|c|}
\hline
& & $\gamma=1.2$ & $\gamma=1.7$ \\
\hline
               & SCDM & $2.03\pm0.015$ & $1.78\pm0.004$\\
$T_m=10,000$ K & OCDM & $2.29\pm0.036$ & $1.81\pm0.012$\\ 
               & LCDM1 & $1.93\pm0.017$ & $1.63\pm0.003$\\
	       & LCDM2 & $2.19\pm0.029$ & $1.83\pm0.008$\\
\hline
               & SCDM & $1.94\pm0.003$ & $1.68\pm0.015$\\
$T_m=60,000$ K & OCDM & $2.10\pm0.014$ & $1.62\pm0.011$\\ 
               & LCDM1 & $1.83\pm0.005$ & $1.53\pm0.012$\\
	       & LCDM2 & $2.08\pm0.013$ & $1.71\pm0.009$\\
\hline
\end{tabular}
\label{los_v}
\end{table}

The dependence of $p$ on the IGM parameters can be understood easily. 
Higher value of $T_m$ implies a larger $x_b$ which, in turn, implies 
more smoothing of the power spectrum at low scales. 
However, the larger scales are more or less 
unaffected by $x_b$. Consequently, the correlation curve becomes flatter as 
we increase $T_m$. Also, for a fixed value of $\gamma$, the 
effect of the cosmological models on the shape of $\xi_{\rm LOS}$  
is large for low $T_m$. 

\begin{figure*}
\psfig{figure=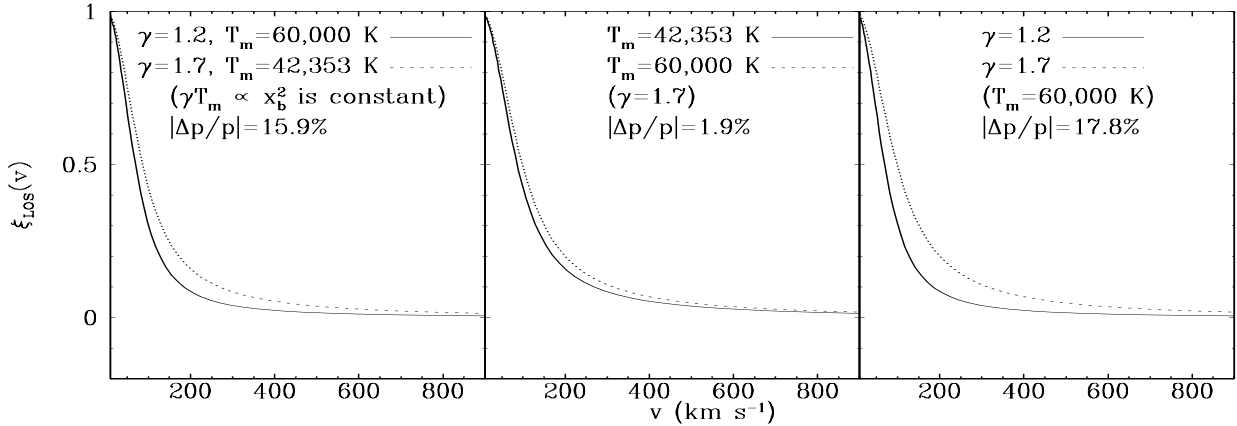,width=18cm}
\caption{
The comparison of effect of $\gamma$ and $T_m$ on the power law index of 
$\xi_{\rm LOS}(v)$. The plots are for LCDM1 model. We give the relative change 
in the power law index $|\Delta p/p|$ for the two curves in each figure.
In the left figure, 
$x_b$ is kept constant. It shows the effect of changing the neutral hydrogen 
density fluctuations. The middle figure shows the effect of changing $x_b$ 
without changing the neutral hydrogen 
density fluctuations (i.e., $\gamma$ is constant). The right figure shows the 
effect of $\gamma$, which affects both $x_b$ and  the neutral hydrogen 
density fluctuations.}
\label{xb_gamma_tm}
\end{figure*}

The effect of $\gamma$ on 
$\xi_{\rm LOS}$ is twofold -- increasing $\gamma$ introduces more 
smoothing at low scales just like $T_m$, and there is also a reduction in the 
neutral hydrogen density fluctuations for given baryonic fluctuations 
see (equation (\ref{eq:nh})). 
Both these effects make the 
correlation curve flatter, which is what we see from Table \ref{los_v}. 
Furthermore, because of this twofold effect, $\gamma$ affects $p$ much more 
than $T_m$ does. This point can be seen clearly in Figure \ref{xb_gamma_tm}. 
In the left figure, we have kept $x_b$ constant, and shown the effect of 
changing only the neutral hydrogen density fluctuations. The middle figure 
shows the effect of changing only $T_m$ or, equivalently, 
the Jeans scale $x_b$, without changing anything else. In the right figure, 
the full effect of $\gamma$ can be seen, as it changes both the Jeans scale 
and the neutral hydrogen density fluctuations.
Since $T_m$ does not have much effect on $p$, we can determine $\gamma$ from 
observations of LOS correlation even with ill-constrained values of $T_m$, 
provided the cosmology is known from some other studies.

Just like in the case of $T_m$, the 
effect of the cosmological models on $p$ 
is large for low $\gamma$. The first reason for this is same as in the case 
of $T_m$ -- low $\gamma$ implies less smoothing and hence DM fluctuations 
are more effective. The second reason is that for low $\gamma$, 
the neutral hydrogen density fluctuations are larger for given 
linear baryonic 
density fluctuations. Thus, a slight 
change in the DM fluctuations causes a significant change in the 
neutral hydrogen fluctuations (for a fixed $x_b$)
which, in turn, makes the correlation function sensitive 
to the DM power spectrum.

As the universe evolves after the 
reionisation, it is possible that 
the value of $\gamma$ increases (Hui \& Gnedin 1997) and 
the LOS correlation becomes less and less 
sensitive to the DM power spectrum. Hence, if the reionisation has occurred 
very early, it is extremely unlikely that one can fix the cosmological 
parameters from observations of $\xi_{\rm LOS}$. We not only need to know 
the value of $\gamma$ and $T_m$ accurately, but also the value of $p$ 
to an accuracy better than 10 per cent.

Finally, we comment on the effect of $h$ on the shape of $\xi_{\rm LOS}$. As 
one can clearly see from Table \ref{los_v}, the effect is quite significant 
($\sim 12$ per cent). The parameter $h$ affects the LOS correlation function 
(equation (\ref{eq:q_los_delz})) in three ways -- (i) it changes the shape 
of $P_{\rm DM}(k)$ (see equation (\ref{eq:ebw})), (ii) it affects the value of 
$x_b$ 
(equation (\ref{eq:xb})) and (iii) it affects the relation between distance 
and redshift. As a result, the values of $\sigma_x(\bar{z})$ and 
$d_H(\bar{z})$ in the integrand of 
equation (\ref{eq:q_los_delz}) get modified. 
However, it turns out that the last two effects
can be scaled out. 
To understand this more clearly, we rewrite 
equation (\ref{eq:q_los_delz}) in a slightly modified form 
\begin{eqnarray}
\lefteqn{
Q_{\rm LOS}(\bar{z},\Delta z)=
\f{D(\bar{z}+\Delta z/2) D(\bar{z}-\Delta z/2)}{2 \pi^2} \times
} \nonumber \\
& & \int_0^{\infty} {\rm d}K 
\left\{\f{\sin (K D_H(\bar{z}) \Delta z)}{K D_H(\bar{z}) \Delta z} \times
\right.
 \nonumber\\
& &\left .\f{h^3 P_{\rm DM}(Kh) K^2{\rm e}^{-K^2 \Sigma_x^2(\bar{z})/2}} 
{(1+X_b^2(\bar{z}+\Delta z/2)K^2)(1+X_b^2(\bar{z}-\Delta z/2)K^2)}\right\}
\nonumber\\
& &
\end{eqnarray}
where $K \equiv k/h$ and 
\be
D_H(\bar{z}) \equiv d_H(\bar{z}) h,~
\Sigma_x(\bar{z}) \equiv \sigma_x(\bar{z}) h,~
X_b \equiv x_b h
\e
One can easily verify that all the three quantities defined above 
$(D_H,\Sigma_x,X_b)$ are 
independent of $h$. Hence $Q_{\rm LOS}(\bar{z},\Delta z)$ 
depends on $h$ only through the combination $h^3 P_{\rm DM}(Kh)$. 
It is very difficult to study this function analytically. 
We have studied it using numerical methods and found that 
for the power spectra used in this paper, its effect 
is to make the LOS correlation steeper when $h$ is 
larger.

\begin{figure*}
\psfig{figure=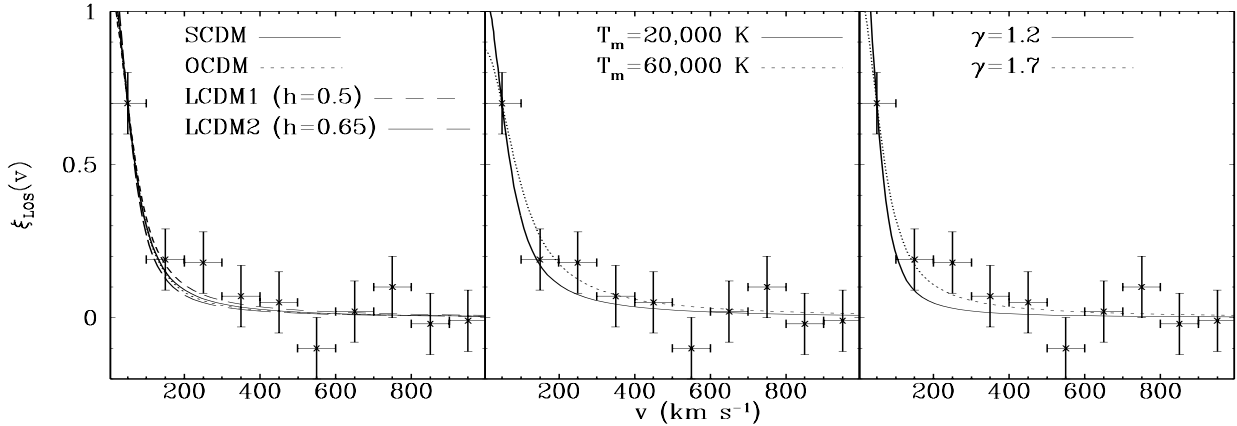,width=18cm}
\caption{
Comparison of the theoretical $\xi_{\rm LOS}$ with observational data 
(Cristiani et al. 1997). The
theoretical curves have been normalised in such a way that they match with the
observed data point at the lowest velocity bin. In the left figure, the IGM 
parameters are fixed to be $T_m=20,000$ K and $\gamma=1.7$. In the 
middle figure, the cosmological model is LCDM1, and $\gamma=1.7$. In the 
right figure, again the cosmology is LCDM1, and $T_m=20,000$ K.} 
\label{chris}
\end{figure*}

To compare our results with observational data we 
take the 
data from Cristiani et al. (1997). The data 
consists of several QSO spectra at various redshifts, ranging from 1.7 to 3.7. 
This range is pretty large, and evolutionary effects will be significant in 
the data. We compare the observed LOS correlation (points with error bars) 
with the theoretical curve for the four cosmological models, and for 
various ranges
of values of $T_m$ and  $\gamma$ in Figure \ref{chris} for $\bar{z}=2.5$.
It should be noted that the observational data points were
obtained using the Ly$\alpha$ clouds with log($N_{\rm HI}/$cm$^{-2}$)$> 14$.
However, we have not used any such constraint while obtaining the
analytical curves.
As a preliminary check, we can see that the analytical curves have the
broad features which are expected from the observational data. We hope to 
carry out a more detailed comparison with observations in a 
future publication.

We next check the redshift evolution of the LOS correlation function.
For definiteness, we consider $\xi_{\rm LOS}$ at a particular velocity,
$v=100$ km s$^{-1}$, and study it as a function of $\bar{z}$. 
We have assumed that $\gamma$ does not
evolve with redshift. Since we are studying the evolution over a large 
redshift range, we have to consider both the possibilities for the 
evolution of $T_m$ discussed in section \ref{mod_par}. 
The evolution 
curve closely follows a power-law dependence, i.e., $\xi_{\rm LOS}(\bar{z}) 
\propto (1+\bar{z})^{-q}$ in the range $1.5<\bar{z}<4.5$. 
We give the values of $q$ and the rms error for different cases 
in Table \ref{los_z}.
\begin{table}
\caption{Power law index, $q$, for the evolution of 
LOS correlation function at velocity separation 100 km s$^{-1}$, defined by 
$\xi_{\rm LOS}(\bar{z}) \propto (1+\bar{z})^{-q}$ in the redshift range 
1.5--4.5. $1\sigma$ errors are given.}
\begin{tabular}{|l|l|c|c|}
\hline
\multicolumn{4}{|c|}{$T_m \sim (1+z)^{3 \gamma-3}$}\\
\hline
& & $\gamma=1.2$ & $\gamma=1.7$ \\
\hline
               & SCDM & $4.00\pm0.045$ & $3.12\pm0.025$\\
$T_m(\bar{z}=2.5)$ & OCDM & $5.69\pm0.066$ & $4.14\pm0.039$\\ 
$=10,000$ K           & LCDM1 & $4.69\pm0.062$ & $3.56\pm0.035$\\
	       & LCDM2 & $5.28\pm0.073$ & $3.91\pm0.042$\\
\hline
               & SCDM & $3.83\pm0.042$ & $3.38\pm0.024$\\
$T_m(\bar{z}=2.5)$ & OCDM & $5.11\pm0.057$ & $4.18\pm0.039$\\ 
$=60,000$ K           & LCDM1 & $4.44\pm0.056$ & $3.73\pm0.034$\\
	       & LCDM2 & $4.96\pm0.067$ & $4.08\pm0.041$\\
\hline \hline
\multicolumn{4}{|c|}{$T_m=$constant}\\
\hline
& & $\gamma=1.2$ & $\gamma=1.7$ \\
\hline
               & SCDM & $3.97\pm0.045$ & $2.98\pm0.026$\\
$T_m=10,000$ K & OCDM & $5.63\pm0.064$ & $3.89\pm0.037$\\ 
               & LCDM1 & $4.66\pm0.061$ & $3.42\pm0.035$\\
	       & LCDM2 & $5.25\pm0.072$ & $3.75\pm0.042$\\
\hline
               & SCDM & $3.65\pm0.039$ & $2.72\pm0.021$\\
$T_m=60,000$ K & OCDM & $4.80\pm0.049$ & $3.24\pm0.025$\\ 
               & LCDM1 & $4.25\pm0.052$ & $3.09\pm0.028$\\
	       & LCDM2 & $4.73\pm0.061$ & $3.33\pm0.033$\\
\hline
\end{tabular}
\label{los_z}
\end{table}

It is clear that the cosmology has maximum influence
when $\gamma$ and $T_m$ 
are small. The reason for this is same as that discussed in the case of 
Table \ref{los_v}.

For a given cosmology and a given value of $\gamma$, the effect of $T_m$ 
(or, equivalently, $x_b$) 
on $q$ is insignificant with the relative change in $q$ being 
about 6--10 per cent. The 
reason for this is as follows:  
the effect of $x_b$ 
is significant only for scales $< x_b$. For the parameters we are considering, 
the velocity scale corresponding to $x_b$ is $\sim$ 10--30 km s$^{-1}$. Hence, 
for the case where $v=100$ km s$^{-1}$, the evolution will not be 
affected significantly by the Jeans length. We studied the evolution at 
a higher velocity scale ($v=250$ km s$^{-1}$), and found that 
the effect of $x_b$ was 
even less (the relative change in $q$ was about 3 per cent at 
$v=250$ km s$^{-1}$).

The effect of $\gamma$ is threefold here -- it affects the value of $x_b$, 
the evolution of $x_b$ (more precisely, this also depends on whether we 
evolve $T_m$ or not) and 
the neutral hydrogen density fluctuations (through the value of $\beta$). 
We have already seen that  
the effect of changing the value of $x_b$ is not very important. 
Increasing the value of $\gamma$ will make the evolution 
of $x_b$ more rapid.  Since $x_b$ appears in the 
denominator of the integrand in equation (\ref{eq:q_los_delz}), the higher 
the value of $\gamma$, the more rapid is the decrease of $\xi_{\rm LOS}$ 
with increasing redshift.
This feature alone will increase 
the value of $q$ with increasing $\gamma$. But, actually
$q$ decreases when we increase $\gamma$ 
because of the third effect of $\gamma$ -- increasing $\gamma$ reduces 
the value
of $\beta$ (equation (\ref{eq:beta})), and $\beta$ 
appears in the exponential in equation (\ref{eq:xi_q}). Hence, 
the $\xi_{\rm LOS}(\bar{z})$ 
curves will decrease less rapidly when $\beta$ is small, i.e., 
$\gamma$ is large.

Finally, we note that the effect of whether $T_m$ evolves or not 
becomes appreciable for large values of $\gamma$. This is obvious because 
larger the value of $\gamma$, more rapid is the evolution of $T_m$. 
Furthermore, we have already argued that if the evolution of $T_m$ is 
more rapid, then the value of $q$ should increase (provided, of course, 
the value of $\beta$ remains unchanged). 
Hence, the values of $q$ are smaller when $T_m$ is kept constant than when 
$T_m$ has a redshift dependence. This can be verified from Table \ref{los_z}.

\begin{figure*}
\psfig{figure=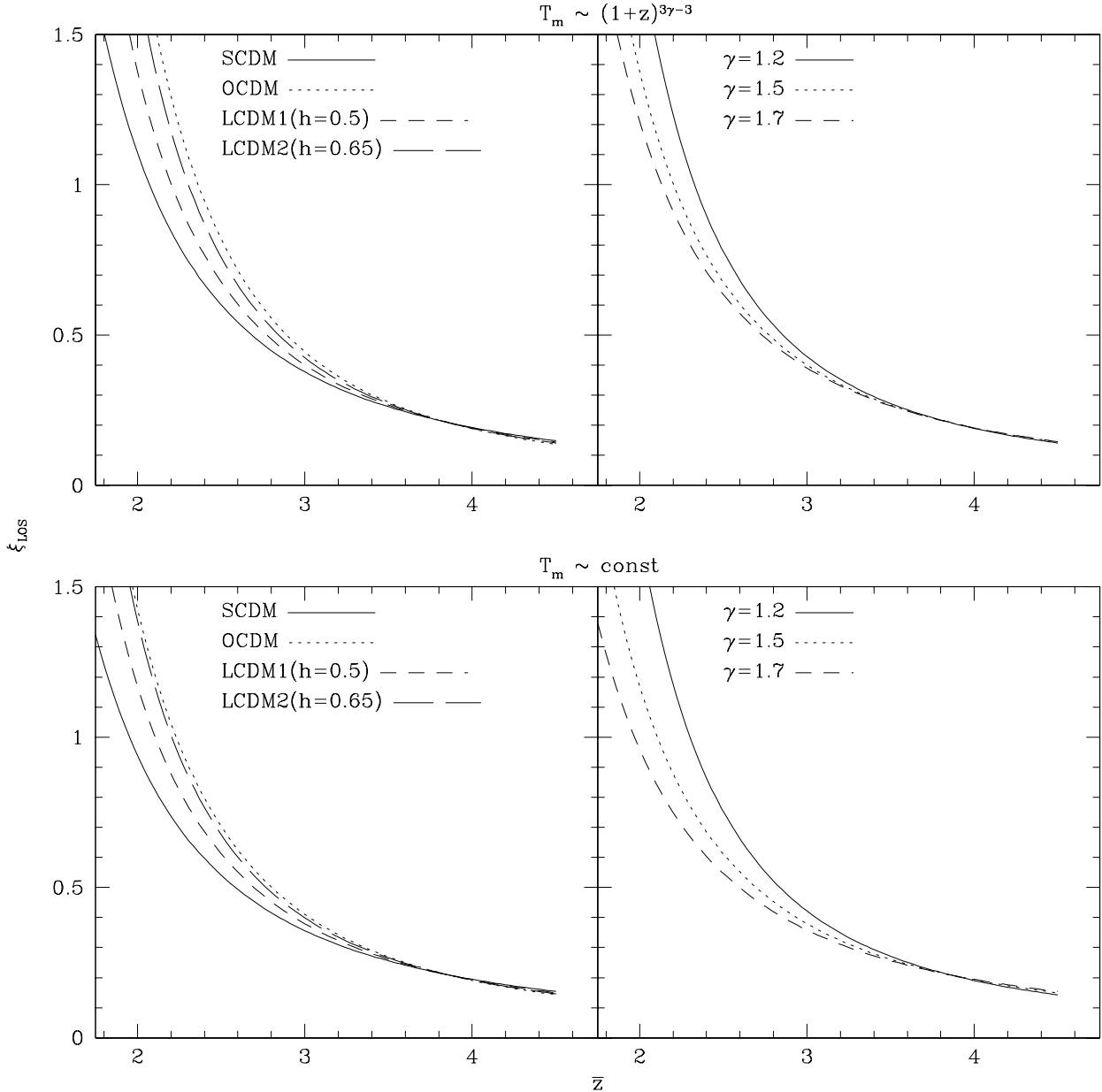,width=18cm}
\caption[]{\label{los_evol}
$\xi_{\rm LOS}$(v=100 km s$^{-1}$) as a function of $\bar{z}$. 
The curves are 
normalised in such a way that $\xi_{\rm LOS}$(v=100 km s$^{-1}$) = 0.21 at 
$\bar{z}=3.85$, which is taken from Cristiani et al.(1997). In the left plots, 
the IGM parameters are fixed $T_m=40,000$ K, $\gamma=1.5$. In the 
right plots, 
we have fixed the cosmological model to be LCDM1, and $T_m=40,000$ K. 
}
\end{figure*}

These effects are shown in Figure \ref{los_evol}. 
The curves are normalised in such a way that
$\xi_{\rm LOS}(v=100$ km s$^{-1}$) = 0.21 at $\bar{z}=3.85$, which is taken
from Cristiani et al.(1997).  
In the top row of Figure
\ref{los_evol}, the value of $T_m$ is fixed at a particular
redshift (in this case, at $\bar{z}=2.5$) and the value of $T_m$ at
other redshifts are calculated using the relation $T_m \sim
(1+z)^{3\gamma -3}$. In the bottom row, $T_m$ is kept 
constant. We do not plot the effect of $T_m$ because we find it to be very 
weak.

The point to be noted here is that 
{\sl the effects due to change in cosmology and $\gamma$ are 
of the same order, which means that we cannot constrain both the 
parameters simultaneously.} Thus the above analysis clearly suggest that 
it will be very difficult to recover the power spectrum of density
fluctuations uniquely from the Ly$\alpha$ absorption lines without knowing
the IGM parameters. 
However, 
the cosmological parameters determined through  
studies such as CMBR (and other data), 
can be used to constrain the equation of state (using the plot in the 
centre in Figure \ref{los_evol}), provided we have some idea about 
the evolution of the Jeans length or equivalently, $T_m$.

Throughout this paper we have treated $\gamma$ to be independent of $z$.
However, there are indications that $\gamma$ could change with $z$.
Schaye et al. (1999b) notice that the temperature of the IGM has a peak
at $z \simeq 3$ and it decreases with decreasing redshift afterwards.
They also notice that the slope of the equation of state become close to one
at $z \simeq 3$ then increases with decreasing redshift. 
Theoretical calculations
suggest that $\gamma$ increases with time and the rate of evolution
depends on the reionisation epoch (Hui \& Gnedin 1997). 
From Table \ref{los_z} and Figure \ref{los_evol}
we can infer that when $\gamma$ becomes larger  the rate of growth of 
$\xi_{\rm LOS}$ at a given velocity decreases.  
{\sl Thus our study clearly suggests
that the evolution of $\xi_{\rm LOS}$ at a given 
velocity can be used as probe of a reionisation and thermal history of the IGM
once the cosmological model and the evolution of $x_b$ is fixed.} We hope 
to study this in detail in a future publication.

\subsection{Transverse Correlation}
\label{sub_trans}

In this section we present the results for the transverse correlation.
As before, we consider the same CDM power spectrum, and essentially the 
same range of the IGM parameters. The smoothing velocity is taken to be 
30 km s$^{-1}$, which 
is the typical peculiar velocity of a blob in the IGM. 

Given $z$, we calculate $\xi_{\rm trans}$ as a function of 
the transverse comoving 
distance $l_{\perp}$. One can then convert this length scale to an 
angular scale 
$\theta$ through the following relations.
\be
\theta=\f{l_{\perp}}{d^{com}_a(z)}
\label{eq:theta}
\e
\be
d^{com}_a(z)=\f{c}{H_0 \sqrt{|\Omega_k|}}\  
S_k \left(x(z)\f{H_0}{c}\sqrt{|\Omega_k|}\right)
\e
where $\Omega_k$ is given by equation (\ref{eq:om_k}), 
$x(z)$ is given by equation (\ref{eq:x(z)}) and
\be
S_k(r)=\left\{ \begin{array}{ll}
	      \sin r &\mbox{(if $\Omega_k < 0$)}\\
	      r &\mbox{(if $\Omega_k = 0$)}\\
	      \sinh r &\mbox{(if $\Omega_k > 0$)}
		\end{array} \right.
\e

\begin{figure*}
\psfig{figure=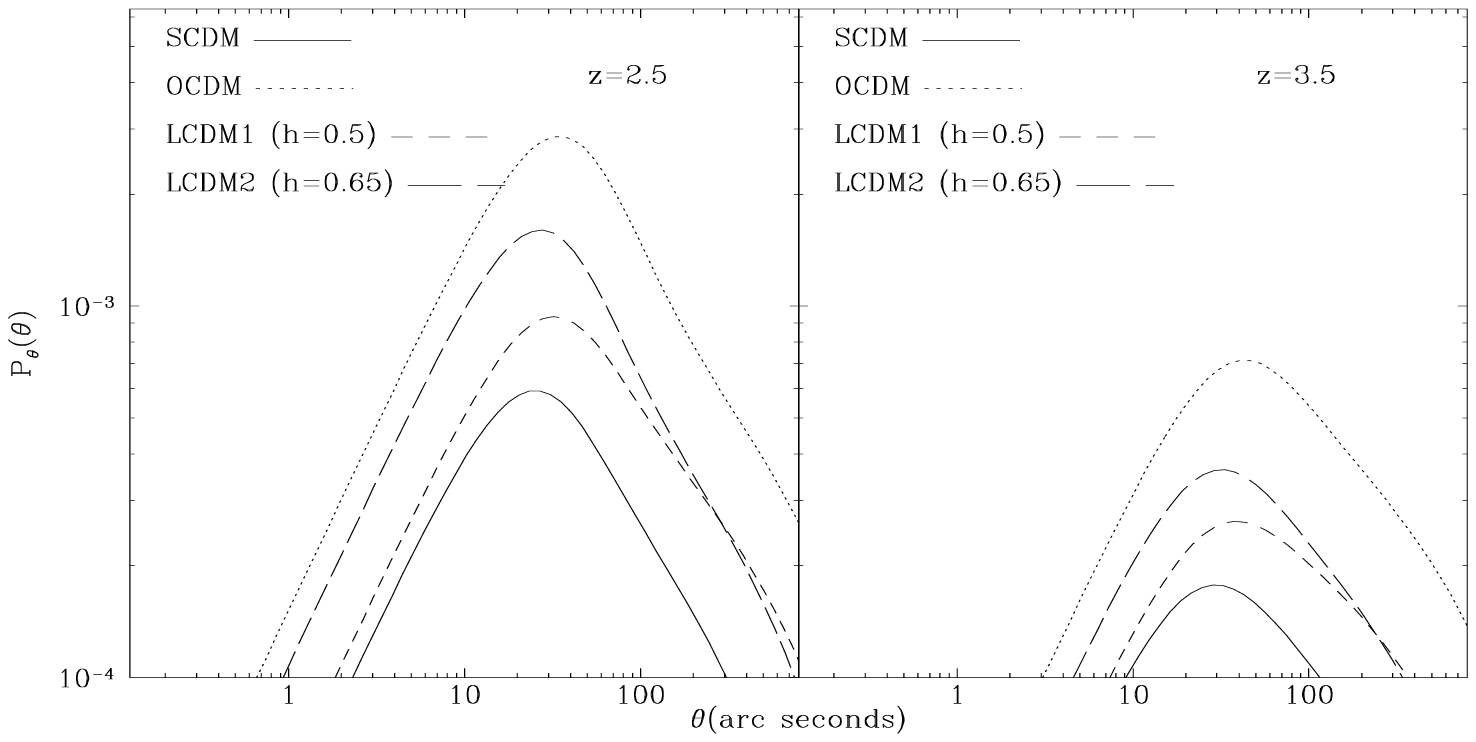,width=18cm}
\caption[]{\label{prob}
Plot of ${\mathcal P}_{\theta}(\theta)$. Results for 
four different cosmological models and for two different redshifts are 
presented. The IGM parameters are $\gamma=1.5, T_m(z=2.5)=40,000$K. $T_m$ at 
redshift $z=3.5$ is calculated using the relation 
$T_m \sim (1+z)^{3\gamma -3}$.}
\end{figure*}

Instead of plotting the correlation function directly, we plot the quantity 
${\mathcal P}_{\theta}(\theta)$, 
which is defined as follows. The excess probability, 
over random background,
of finding two neutral hydrogen overdense regions 
separated by a comoving transverse 
distance $l_{\perp}$ is
\be
{\mathcal P}_{l_{\perp}}(l_{\perp}) {\rm d}l_{\perp}=\f
	{\xi_{\rm trans}(l_{\perp};z) 2 \pi l_{\perp} {\rm d}l_{\perp}}
	{4 \pi [d^{com}_a(z)]^2}.
\e
Using equation (\ref{eq:theta}) we get the excess probability 
over random background
of finding two neutral hydrogen overdense regions 
separated by an angle $\theta$ as
\begin{eqnarray}
{\mathcal P}_{\theta}(\theta)~{\rm d}\theta &=&
{\mathcal P}_{l_{\perp}}(l_{\perp}) \frac{{\rm d}l_{\perp}} {{\rm d}\theta} 
{\rm d}\theta
\nonumber \\
& = & \f{1}{2}\xi_{\rm trans}(\theta)~\theta~{\rm d} \theta.
\end{eqnarray}
From Figure \ref{prob} it is clear that even the maximum excess probability 
of finding two H~{\sc i} overdense regions over an angular scale 
greater than few arc seconds is less than 1 per cent. Observationally the
distribution of H~{\sc i} along the transverse direction is 
probed by studying the common absorbers along the LOS 
towards closely spaced QSOs. The angular scales probed varies
between few arc seconds and few arc minutes 
(Shaver, Boksenberg \& Robertson 1982; Shaver 
\& Robertson 1983; Smette et al. 1992; Dinshaw et al. 1994; 
Bechtold et al. 1994; Crotts et al. 1994; Bechtold \& Yee 1994; 
Smette et al. 1995; D'Odorico et al. 1997; Petitjean et al. 1998). 
Based on our
analysis it is most likely that the common absorbers seen
in the spectra of closely spaced QSOs are most likely probe
the transverse extent of the same overdense region rather than
the clustering length scale of separate regions.

\subsection{Difference between $\xi_{\rm trans}$ and $\xi_{\rm LOS}$}
\label{sub_rat}

It should be noted that for a given mean redshift, the values of the LOS and
the transverse correlation functions need not be the same. This is because, 
when we
observe along one LOS, we actually sample different points at different
redshifts. In contrast to this, the transverse correlation is calculated 
at the same
redshift. The effect of evolution in the LOS correlation makes it different
from the transverse correlation.

To illustrate this point more clearly, let us first assume that $x_b$ does not 
evolve with $z$. Then from equations (\ref{eq:q_los}) and (\ref{eq:q_sp}), we 
see that for a given length scale $l$, the integrands in the two equations 
are identical. Hence, we get
\be
\f{Q_{\rm LOS}}{Q_{\rm trans}}=
\f{D(\bar{z}+\Delta z/2)D(\bar{z}-\Delta z/2)}
                    {D^2(\bar{z})},
\e
where $\bar{z}$ and $\Delta z$ are defined in equations (\ref{eq:z_bar}) and 
(\ref{eq:z_dif}) respectively. The difference in the two correlation functions 
is now entirely due to the evolution of the power spectrum. 
Thus the two correlation functions will be nearly equal for small $\Delta z$ 
but will start differing from each other for large 
$\Delta z$. In the general case when $x_b$ evolves with $z$, the difference 
will be much more prominent.
\begin{figure*}
\psfig{figure=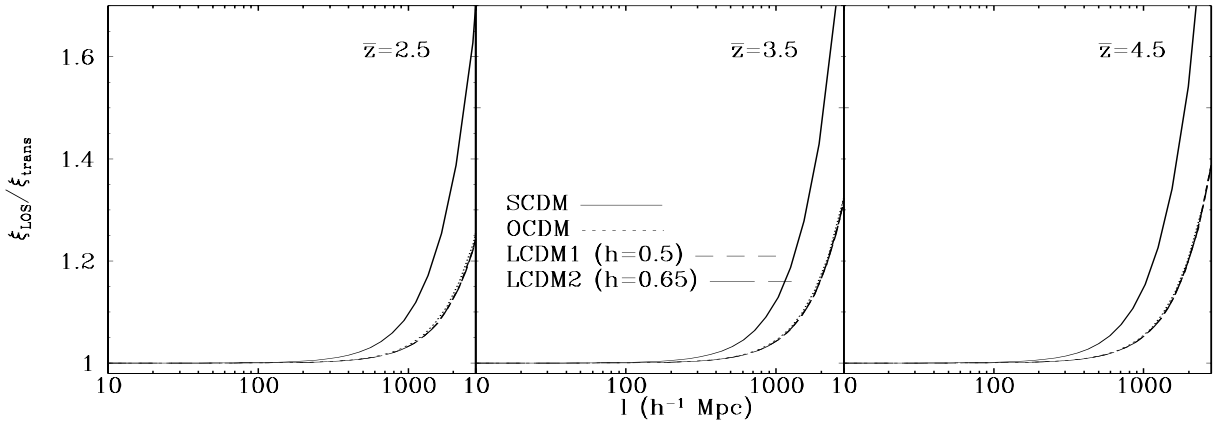,width=18cm}
\caption[]{\label{rat}
The ratio of $\xi_{\rm LOS}$ and $\xi_{\rm trans}$ 
as a function of comoving scale $l$.
The results for four cosmological models and for three different mean 
redshifts are plotted. The curves for the LCDM1, LCDM2 
and the OCDM models nearly overlap. The model parameters relating to the IGM 
are $\gamma=1.5, T_m(\bar{z}=2.5)=40,000$K. $T_m$ at the other two 
redshifts are calculated using the relation 
$T_m \sim (1+z)^{3\gamma -3}$.}
\end{figure*} 

This is indeed true, as one can see from Figure \ref{rat}. For scales below
200 \hmpc, the two correlation functions are nearly the same. But above
such scales the two functions start differing appreciably. For the
observations made in the scales of 10--100 \hmpc, our analytical
calculation shows that one should not see any appreciable difference between
LOS and transverse correlations. This can be used as a important tool 
determining the 
power spectrum (provided, of course, we know 
the IGM parameters and the correlation function completely). 
As we have argued earlier, one cannot get the power spectrum 
from the LOS correlation. But the power spectrum can be obtained from the 
transverse 
correlation in usual manner. Since the two correlations are identical for 
scales upto 100 \hmpc, one can start from the LOS correlation, replace it with 
the transverse correlation, and obtain the power spectrum. 

\subsection{Column Density Distribution}
\label{sub_nhres}

\begin{figure}
\psfig{figure=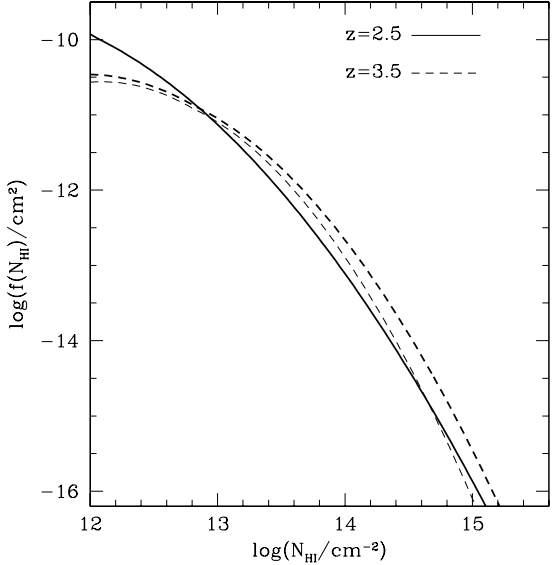,width=8cm}
\caption[]{\label{nh_comp}
Column density distribution of neutral hydrogen plotted for different kinds 
of evolution of $T_m$ and $T_0$ for two redshifts. 
The thicker lines represent the case where 
both of them are constant. For definiteness we have taken $T_m=40,000$ K and 
$T_0=20,000$ K. The thinner lines are for the case where both the temperatures 
vary as $(1+z)^{3 \gamma-3}$. The values are $T_m(z=2.5)=40,000$ K and 
$T_0(z=2.5)=20,000$ K. Obviously, the curves for the two cases overlap at 
$z=2.5$. The cosmology is taken to be LCDM1, $\gamma=1.7$ and $\epsilon=0.3$.}
\end{figure}

In this section we study the results for the column density distribution. 
As we have discussed in section \ref{mod_par}, 
we shall consider two separate cases for 
the evolution of $T_0$ and $T_m$, i.e., 
(i) $(T_0$ and $T_m) \propto (1+z)^{3 \gamma -3}$ 
and (ii) $T_0$ and $T_m$ are  
independent of $z$. The comparison between the two cases is shown in 
Figure \ref{nh_comp}, where we plot the quantity $f(N_{\rm HI})$ for 
the two different cases. $f(N_{\rm HI})$ is 
related to
$({\rm d}N_{\rm{pk}}/{\rm d}z~{\rm d}N_{\rm HI})$ through the relation 
(Bi \& Davidsen, 1997)
\be
f(N_{\rm HI})=({\rm d}N_{\rm{pk}}/{\rm d}z~{\rm d}N_{\rm HI})/(1+z).
\e

As we can see from Figure \ref{nh_comp}, 
the difference between the two cases becomes more significant 
at higher column densities. We have checked and found 
that $T_m$ has very little effect on the column density distribution. The 
difference between the two cases at $z=3.5$ 
is because of the fact that 
the value of $T_0$ is different for the two cases at 
that redshift. However, we have already mentioned that any uncertainty 
in the knowledge of 
$T_0$ can be compensated to some extent by changing $\epsilon$. 
Hence, for studying the column density distribution 
we shall consider only the case 
where both $T_m$ and $T_0$ evolve as $(1+z)^{3\gamma-3}$.

We shall now discuss the dependence of the 
column density distribution on the following parameters : (i) cosmological 
models, (ii) $\epsilon$ and (iii) $\gamma$. 

\begin{figure*}
\psfig{figure=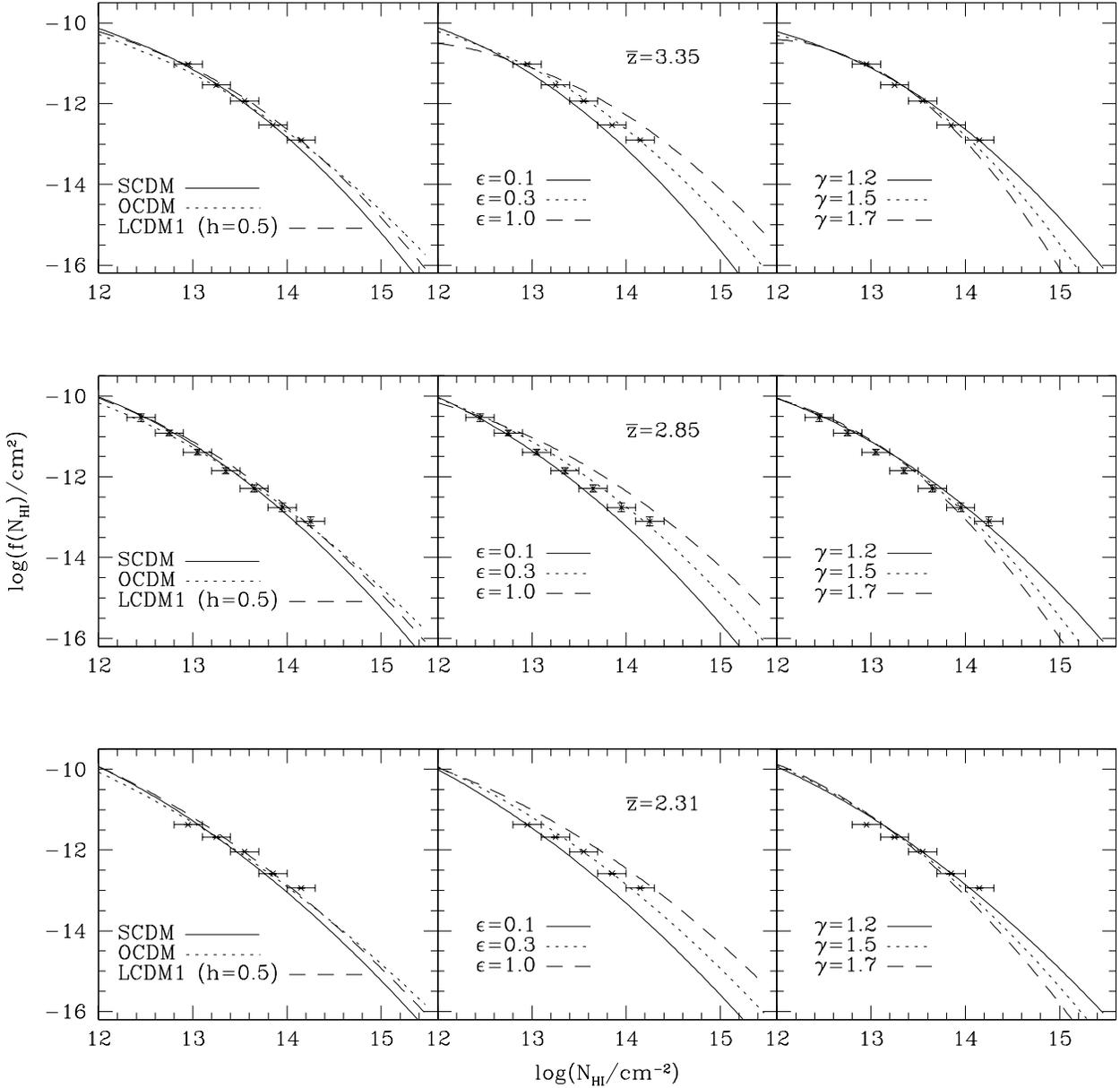,width=18cm}
\caption[]{\label{nh}
$f(N_{\rm HI})$ as a function of $N_{\rm HI}$ for three redshifts. 
We show the dependence of $f(N_{\rm HI})$ on 
cosmology (left), $\epsilon$ (centre) and equation of state 
$\gamma$ (right). For all the plots we have taken $T_m(z=2.5)=40,000$ K and 
$T_0(z=2.5)=20,000$ K. For the left plots, we fix $\gamma=1.2, \epsilon=0.3$. 
We have not plotted the curve for the LCDM2 model because it completely 
overlaps with that for the LCDM1 model.
The centre plots are for the LCDM1 model with 
$\gamma=1.2$. The right plots are also for LCDM1 with $\epsilon=0.3$.
}
\end{figure*}

In what follows we try to get constraints on our model
parameters using the observed column density distribution 
obtained from Hu et al. (1995) and Kim et al. (1997) at three mean 
redshifts 2.31, 2.85 and 3.35.
In Figure \ref{nh} we have plotted the quantity $f(N_{\rm HI})$ for 
those redshifts.
The observational data points are the points with errorbars in the figure.  

For all the plots in Figure \ref{nh}, we have fixed $T_m(z=2.5)=40,000$ K and 
$T_0(z=2.5)=20,000$ K.
In the  left most panel of each row in the figure  
we plot the predicted column density distribution for various 
cosmological models for a given set of IGM parameters 
($\gamma =1.2$) and $\epsilon= 0.3$. We did not plot the LCDM2 model because 
it completely overlaps with the LCDM1 model.
In all the redshift bins it 
is clear that the SCDM curves fall steeply at the higher column 
density end compared to other models. This is consistent with
the results noted by Gnedin \& Hui (1996). However, in our method of 
obtaining $f(N_{\rm HI})$, 
the SCDM model can be made to fit the data by slightly
increasing the value of $\epsilon$. 
The OCDM and LCDM1 curves fit the
observed distribution upto $\log(N_{\rm HI}/{\rm cm}^{-2}) = 14.2$ 
for $\epsilon = 0.3$. 

In the
middle panel of each row in Figure \ref{nh} we plot the predicted 
distribution
for the three assumed values of $\epsilon$ for LCDM1 with 
$\gamma = 1.2$. Increasing the value of $\epsilon$ enhances the column 
density of a cloud (with fixed peak density), and consequently, the 
value of $f(N_{\rm HI})$ 
increases at the high column density region.
It is clear that the observed distribution is consistent
with $\epsilon = 0.3$. 
This means that, in the case of LCDM,  the effective length
of the overdense region is about (1/3) of  the coherence scale $R^*$. 
The value of $R^*$ depends on the baryonic 
power spectrum, its typical value is of the order of few hundreds of Kpc; for 
LCDM model with $\gamma=1.2, T_m(z=2.5)=40,000$K, 
we get $R^* \simeq 60$\hkpc. Hence 
the typical length of an individual overdense region is $\sim 20$\hkpc. 
Zhang et al. (1998) infer that the typical LOS scale length 
of an absorbing cloud 
is $\sim$ 15--35 \hkpc\, through their hydrodynamical simulations. Our 
values are consistent with theirs.

The effect of $\gamma$ on the column density distribution can be seen
from the right most panel in Figure \ref{nh} where we plot the results
of LCDM1 model for three values of $\gamma$. In the
models with higher value of $\gamma$, the $f(N_{\rm HI})$ curve falls sharply
at higher column density end at all redshifts, which is 
consistent with the conclusions of Hui et al. (1997). The reason for this 
sharp fall is because the 
neutral hydrogen density is less (for a given baryonic density) 
when the value of $\gamma$ is more and, hence 
there are less number of clouds with high column densities.
At low redshifts, it is clear that lower values of 
$\gamma$ are preferred for $\epsilon=0.3$. 
However, one can get the consistent fit for higher values of $\gamma$ 
by increasing the value of $\epsilon$.
For example, at $\bar{z}=2.31$, one can fit the data reasonably well 
with $\epsilon$ in the range 0.3 to 0.5 
and $\gamma$ in the range 1.2 to 1.6. But 
it is impossible to fit the data with 
$\gamma > 1.6$, for any choice of $\epsilon$ at $\bar{z}=2.31$. 
This implies that we cannot accommodate 
$\gamma$ larger than 1.6 at this redshift even 
by changing the values of $T_0$ or other IGM parameters.
At $\bar{z}=2.85$, one can fit the observations 
for $\gamma =1.7$ (the highest value of $\gamma$ we are considering 
in this paper), 
by taking $\epsilon=0.4$. For the 
case where $\bar{z}=3.35$, we can see that we can marginally 
fit the observations for whole range of $\gamma$ considered in this paper 
with $\epsilon=0.3$ itself.

Finally, we comment on how our results compare with some of the 
hydrodynamical simulations. 
It has been observed through numerical 
simulations that the peak baryonic over-density and the neutral hydrogen 
column density are strongly 
correlated (Zhang et al. 1998; Dav\'e et al. 1999). We tested 
equation (\ref{eq:n_b_nh}) against the scatter plots showing the correlations. 
The parameters were 
taken to be: LCDM1 cosmological model with $\gamma=1.2, \epsilon=0.3$ at 
$\bar{z}=3$. These are the parameters which fit the observations for the 
column density distribution 
reasonably well (see Figure \ref{nh}). It turns out that the 
relation matches well with the median of the scatter plot in Figure 8 of 
Zhang et al. (1998). Dav\'e et al. (1999) give an analytical formula 
which relates $n_{\rm B}[{\rm peak}]/n_0$ and $N_{\rm HI}$ by a 
power-law (see equation (7)). 
We find that one needs a much higher value of $\gamma$ ($\sim 1.82$) 
in our model to match 
the power-law index at redshift 3. 
Also, one needs a $\epsilon \sim 1$ to match the overall scaling factor. 
If we take such values of $\epsilon$ and $\gamma$, then our results for 
the column density distribution
will not be able to match the observations. We believe that such a discrepancy 
arises because of the difference in methods used for 
assigning a column density 
to a cloud.

We next test the column density distributions against some of the recent 
simulation results. 
In the column density range $10^{12.8}$--$10^{14.3}$ cm$^{-2}$, one can 
fit a power-law of the form 
$f(N_{\rm HI}) \propto N_{\rm HI}^{-\beta_{\rm HI}}$. The values of 
$\beta_{\rm HI}$ for our LCDM1 model with $\gamma=1.2$ and $\epsilon=0.3$ are 
$\beta_{\rm HI}=1.70\pm0.02$ for $\bar{z}=2.31$, 
$\beta_{\rm HI}=1.65\pm0.04$ for $\bar{z}=2.85$,
$\beta_{\rm HI}=1.56\pm0.04$ for $\bar{z}=3.35$. When compared with 
fits to the observational points, the corresponding power-law indices are 
$1.35\pm0.03,~1.39\pm0.26,~1.59\pm0.13$, respectively (Kim et al. 1997). All 
errors given here are $2 \sigma$. Simulations of 
Zhang et al. (1997) using the SCDM model produce a 
$\beta_{\rm HI}=1.39\pm0.06$ in the range 
$2 \times 10^{12} < N_{\rm HI}/{\rm cm}^{-2} < 10^{14}$ for $z=3$. 
Our curves are also in quite good agreement with the P$^3$M--SPH simulations 
(using SCDM model) by 
Theuns, Leonard \& Efstathiou (1998) and Theuns et al. (1998) 
in the same column density and redshift ranges. 
More recently, Machacek et al. (2000) have performed hydrodynamical 
simulations for various cosmological models. The LCDM model ($\Omega_m=0.4, 
\Omega_{\Lambda}=0.6,h=0.65, \sigma_8=1.0$) 
with $\Omega_{\rm B} h^2=0.015$ gives 
$\beta_{\rm HI}=1.61\pm0.04$ for $z=2$ and  
$\beta_{\rm HI}=1.48\pm0.04$ for $z=3$ in the range 
$10^{12.8} < N_{\rm HI}/{\rm cm}^{-2} < 10^{14.3}$. 
They note that the power-law 
index is higher for the SCDM model, and lower for the OCDM model. This is 
also consistent with what we get from our model.

\section{Conclusions}
\label{concl}

We have presented a simple analytic expression for the correlation
function and the column density distribution for the low H~{\sc i} 
column density systems seen in the
spectra of high redshift QSOs. We have used our results to get constraints 
on various cosmological and IGM parameters. We summarise our main results 
below.

1. One cannot rigourosly define a power spectrum from the LOS correlation 
function. 
However, since the LOS and transverse correlations are identical below 
scales of $\sim 100$ \hmpc, it is possible to obtain the power spectrum by 
a Fourier transform of the LOS correlation function (provided the IGM 
parameters are known). 

Previous studies have attempted to recover the power spectrum of density 
fluctuations from the observations of the IGM (Croft et al. 1998, Hui 1999, 
Croft et al. 1999). We show that 
it is difficult to recover a unique power spectrum from 
H~{\sc i} correlation function 
without constraining the IGM parameters, especially $\gamma$. We have shown 
that the shape of the LOS correlation function at a particular 
mean redshift becomes less and less sensitive to the DM power spectrum 
as the universe evolves after reionisation. We feel that 
the correct approach in studying this issue is to constrain the cosmological 
models using CMBR or supernovae data, and apply those constraints to study 
the IGM parameters using H~{\sc i} correlation functions.

2. The LOS correlation function and its evolution is much more sensitive to 
$\gamma$ than $T_m$. Using observations which give the shape of the LOS 
correlation at a particular redshift, one can constrain the value of 
$\gamma$, even with ill-constrained values of $T_m$, provided the 
background cosmology is known. However, the more uncertain is the 
value of $T_m$, the less constrained is $\gamma$. 
Carrying out such an exercise for 
different redshifts, it will be possible to constrain 
the evolution of $\gamma$. However, for this study, one needs 
accurate observational data at different redshift bins which are not affected 
by evolutionary effects. 
{\it Thus one can use the study of correlation function 
as an independent 
method to constrain the reionisation history of the universe.}

3. The analytic column density distribution for H~{\sc i}, like 
$\xi_{\rm LOS}$, is less sensitive to $T_m$ than $\gamma$. The 
distribution, when compared with 
observations, favours a lower value of $\gamma$, although at 
redshifts $> 2.5$, 
one can marginally fit the observations with higher values of $\gamma$. 
Our model clearly rules out $\gamma>1.6$ at redshift 2.31.

\section*{Acknowledgment}

We gratefully acknowledge the support from the Indo-French Centre for 
Promotion of Advanced  Research under contract No. 1710-1.
TRC is supported by the University Grants Commission, India. We would like to 
thank the anonymous referee for suggestions which helped us to improve 
the clarity of the paper. We also thank Patrick Petitjean for useful comments.

\appendix
\section{Detailed calculations for the column density distribution}

Suppose we are looking at the IGM along any one direction, at some redshift 
$z$. Then the linear 
density field $\delta_{\rm B}^{(1D)}(x,z)$ along 
that axis will be described by a 
one 
dimensional Gaussian 
random field, with a power spectrum 
\be
P_{\rm B}^{(1D)}(k,z)=\f{1}{2\pi} D^2(z)\int_k^{\infty} 
{\rm d}k^{\prime} k^{\prime} \f{P_{\rm DM}(k)}{(1+x_b^2(z)k^2)^2}
\e

From now on we shall derive all the expressions at a particular redshift 
$z$, and we shall not write the explicit $z$-dependence on the quantities.
 
To define the column density, we associate each local maximum or peak in the 
linear density field to a Ly$\alpha$ cloud. The column density corresponding 
to such a cloud is given by equation (\ref{eq:col_den}).
We have expressed $\delta_{\rm B}^{(1D)}$[peak] 
in terms of $N_{\rm HI}$ in
equation (\ref{eq:delta_b_nh}).

Using the properties of a Gaussian random field, we can derive the joint 
probability distribution for the three Gaussian random fields 
$\delta_{\rm B}^{(1D)},
\delta_{\rm B}^{(1D) \prime \prime},\delta_{\rm B}^{(1D) \prime}$. 
The probability that 
the field and its second derivative have values 
$\delta_{\rm B}^{(1D)}$ and $\delta_{\rm B}^{(1D) \prime \prime}$, 
respectively at the 
peak $\delta_{\rm B}^{(1D) \prime}=0$ is
\begin{eqnarray}
\lefteqn{
{\mathcal{P}}
[\delta_{\rm B}^{(1D)},\delta_{\rm B}^{(1D) \prime \prime},
\delta_{\rm B}^{(1D) \prime}=0]~ 
{\rm d}\delta_{\rm B}^{(1D)}~{\rm d}\delta_{\rm B}^{(1D) \prime \prime}~ 
|\delta_{\rm B}^{(1D) \prime \prime}|~{\rm d}x=} \nonumber\\
& &
\f{1}{(2 \pi)^{3/2} \sigma_1 \Sigma} 
\exp \left[\f{1}{2 \Sigma^2} (\sigma_2^2 \delta_{\rm B}^{(1D)^2} 
+2 \sigma_1^2 \delta_{\rm B}^{(1D)}\delta_{\rm B}^{(1D) \prime \prime}
\right.
\nonumber \\
& &\left. + \sigma_0^2 \delta_{\rm B}^{(1D) \prime \prime^2})\right]
{\rm d}\delta_{\rm B}^{(1D)}~{\rm d}\delta_{\rm B}^{(1D) \prime \prime}~ 
|\delta_{\rm B}^{(1D) \prime \prime}|~{\rm d}x,
\end{eqnarray}
where
\be
\sigma_m^2 \equiv \sigma_m^2(z) 
=\f{1}{2\pi}\int_{-\infty}^{\infty}{\rm d}k~k^{2 m}P_{\rm B}^{(1D)}(k,z),
\label{eq:sigma_m}
\e
and
\be
\Sigma^2=\sigma_0^2\sigma_2^2-\sigma_1^4.
\e
Note that $\sigma_0 =\Delta$, defined in equation (\ref{eq:deltasq}).

For convenience, let us define some dimensionless quantities
\be
\nu \equiv \f{\delta_{\rm B}^{(1D)}}{\sigma_0},
\lambda \equiv -\f{\delta_{\rm B}^{(1D) \prime \prime}}{\sigma_2}, 
\kappa \equiv \f{\sigma_1^2}{\sigma_0 \sigma_2}
\label{eq:nu}
\e
$\nu$ and $\lambda$ measure the field and its second derivative, 
respectively; $\kappa$ is a measure of the width of the power spectrum. 
One can use 
these quantities to obtain the number of peaks (clouds) per unit length
\begin{eqnarray}
\lefteqn{
\f{{\rm d}N_{\mbox{pk}}}{{\rm d}x}=} \nonumber \\
& & {\mathcal{P}}
[\delta_{\rm B}^{(1D)},\delta_{\rm B}^{(1D) \prime \prime},
\delta_{\rm B}^{(1D) \prime}=0] 
~{\rm d}\delta_{\rm B}^{(1D)}~{\rm d}\delta_{\rm B}^{(1D) \prime \prime}~ 
|\delta_{\rm B}^{(1D) \prime \prime}|
\end{eqnarray}
Using equation (\ref{eq:delta_x}), one can convert the above expression to 
the number of clouds per unit 
redshift interval. After simplification, the relation 
becomes
\begin{eqnarray}
\lefteqn{
\f{{\rm d}N_{\rm{pk}}}{{\rm d}z} =
\f{d_H(z)}{(2 \pi)^{3/2} \sqrt{1-\kappa^2} R^*} \times
}\nonumber\\
& &
\exp\left[-\f{1}{2}\left\{\f{(\nu-\kappa\lambda)^2}{1-\kappa^2}+\lambda^2
\right\}\right] \lambda~{\rm d}\lambda~{\rm d}\nu
\end{eqnarray}
The $\lambda$ integration can be carried out to obtain
\begin{eqnarray}
\lefteqn{
\f{{\rm d}N_{\mbox{pk}}}{{\rm d}z~{\rm d}\nu}=
\f{d_H(z)}{(2 \pi)^{3/2} R^*}\left[
	\sqrt{1-\kappa^2} \exp\left(-\f{\nu^2}{2(1-\kappa^2)}\right) \right.
}\nonumber \\
& &
+\kappa \nu \sqrt{2 \pi} {\rm e}^{-\nu^2/2} \nonumber \\
& & \left.
    -\sqrt{\f{\pi}{2}} \kappa \nu~ 
	\mbox{erfc}\left(\f{\kappa\nu}{\sqrt{2(1-\kappa^2)}}\right) 
	{\rm e}^{-\nu^2/2}\right]
\label{eq:dndzdnu}
\end{eqnarray}
We are interested in the quantity
\be
\f{{\rm d}N_{\rm{pk}}}{{\rm d}z~{\rm d}N_{\rm HI}} =
\f{{\rm d}N_{\rm{pk}}}{{\rm d}z~{\rm d}\nu} \f{{\rm d}\nu}{{\rm d}N_{\rm HI}}
\e
which is straightforward to obtain from equation (\ref{eq:dndzdnu}), provided 
we know $\nu$ as a function of $N_{\rm HI}$. 
Equations (\ref{eq:delta_b_nh}) and 
(\ref{eq:nu}) give $\nu$ in terms of $N_{\rm HI}$, 
and they can be used to calculate 
${\rm d}\nu/{\rm d}N_{\rm HI}$ (for $n_{\rm HI} \ll n_{\rm B}$)
\be
\f{{\rm d}\nu}{{\rm d}N_{\rm HI}}=\f{1}{\beta \Delta N_{\rm HI}}.
\e
Thus we get an analytic expression for the 
number of clouds per unit redshift interval per unit column density range 
$({\rm d}N_{\rm{pk}}/{\rm d}z~{\rm d}N_{\rm HI})$ 
as a function of $N_{\rm HI}$.

\label{lastpage}

\end{document}